\documentclass[pdflatex,sn-chicago]{sn-jnl}

\jyear{2024}%

\theoremstyle{thmstyleone}%
\newtheorem{lemma}{Lemma}%

\theoremstyle{plain}
\newtheorem{propo}{Proposition}
\newtheorem*{thm*}{Theorem}

\theoremstyle{definition}

\newcommand{\argmin}{\mathop{\rm arg~min}\limits}
\providecommand{\abs}[1]{\lvert#1\rvert}

\usepackage{bm}
\usepackage{etoolbox} 
\usepackage{amsthm}
\usepackage{float}
\usepackage{here}
\usepackage{subcaption}

\makeatletter
\patchcmd{\ps@headings}
{\hbox to \hsize{\hfill Springer Nature 2021 \LaTeX\ template\hfill}}
{\hbox to \hsize{\hfill \hfill}}
{}
{}
\makeatother
\makeatletter
\patchcmd{\ps@headings}
{\hbox to \hsize{\hfill Springer Nature 2021 \LaTeX\ template\hfill}}
{\hbox to \hsize{\hfill \hfill}}
{}
{}
\makeatother
\makeatletter
\patchcmd{\ps@headings}
{\hbox to \hsize{\hfill Springer Nature 2021 \LaTeX\ template\hfill}}
{\hbox to \hsize{\hfill \hfill}}
{}
{}
\makeatother

\begin{document}

\title[Multi-task learning via robust regularized clustering]{Multi-task learning via robust regularized clustering with non-convex group penalties}


\author*[1]{\fnm{Akira} \sur{Okazaki}}\email{okazaki.akira.864@s.kyushu-u.ac.jp}

\author[2]{\fnm{Shuichi} \sur{Kawano}}

\affil[1]{\orgdiv{Graduate School of Mathematics}, \orgname{Kyushu University}, \orgaddress{\street{744 Motooka}, \city{Nishi-ku}, \state{Fukuoka}, \postcode{819-0395} \country{Japan}}}
\affil[2]{\orgdiv{Faculty of Mathematics}, \orgname{Kyushu University}, \orgaddress{\street{744 Motooka}, \city{Nishi-ku}, \state{Fukuoka}, \postcode{819-0395} \country{Japan}}}

\abstract{
Multi-task learning (MTL) aims to improve estimation and prediction performance by sharing common information among related tasks. One natural assumption in MTL is that tasks are classified into clusters based on their characteristics. However, existing MTL methods based on this assumption often ignore outlier tasks that have large task-specific components or no relation to other tasks. To address this issue, we propose a novel MTL method called Multi-Task Learning via Robust Regularized Clustering (MTLRRC). MTLRRC incorporates robust regularization terms inspired by robust convex clustering, which is further extended to handle non-convex and group-sparse penalties. The extension allows MTLRRC to simultaneously perform robust task clustering and outlier task detection. The connection between the extended robust clustering and the multivariate \textit{M}-estimator is also established. This provides an interpretation of the robustness of MTLRRC against outlier tasks. An efficient algorithm based on a modified alternating direction method of multipliers is developed for the estimation of the parameters.
The effectiveness of MTLRRC is demonstrated through simulation studies and application to real data.}




\maketitle

\section{Introduction}\label{sec1}
Multi-task learning (MTL) \citep{Caruana1997-bb} is a methodology where we simultaneously learn multiple related tasks so that each task uses information from other tasks. In the field of statistics, MTL is formulated as an estimation problem for multiple statistical models corresponding to multiple datasets with prior information about relationships among models.
Because, in real problems, related tasks tend to have some common information, MTL is expected to have better estimation and prediction accuracy than independently estimated models \citep{Zhang2021-ia}. The key part of MTL is the assumption of how tasks relate to each other and transfer the common information across models.  When tasks are sufficiently related as assumed, MTL can leverage the shared information to improve the efficiency of estimation and generalization performance. However, when tasks are unrelated or only weakly related, forcing information sharing among them can be detrimental.

 A reasonable assumption to address this issue is that tasks are classified into some groups sharing common characteristics. MTL methods based on this assumption are achieved by clustering the parameters of models. For instance, \cite{Argyriou2007-cj} introduced the $k$-means algorithm for task clustering, while several other studies (\cite{Zhong2012-il}; \cite{Yamada2017-cg}; \cite{He2019-wy}; \cite{Dondelinger2020-wz}; \cite{Okazaki2023-rc}) utilized fused regularization techniques such as fused lasso \citep{Tibshirani2005-nk} and network lasso \citep{Hallac2015-ss}. However, these methods have not considered the existence of outlier tasks that share common characteristics but significantly have task-specific uniqueness or do not share any common characteristics with other tasks.
 The clustering techniques employed for MTL intend to classify all tasks into some clusters and estimate the model to be close to the center of the cluster. Thus, their existence would deteriorate the performance of clustering and lead to a false interpretation of the estimated clusters. On the other hand, some robust MTL methods (\cite{Chen2011-yr}; \cite{Gong2012-gl}) have been proposed to detect them and reduce their influence on a common structure. However, these methods also have certain limitations. First, they typically classify tasks into only one shared structure and other outlier tasks. If multiple common characteristics exist among tasks, it would be difficult to extract true task structures. Second, these methods often rely on the group lasso regularization \citep{Yuan2006-pd} for outlier parameters to identify the outlier tasks, which restricts the value of outlier parameters. Limiting the value of the outlier parameters may ignore the nature of outlier tasks.

To address these limitations of existing MTL methods, we propose a novel robust MTL method called Multi-Task Learning with Robust Regularized Clustering (MTLRRC). MTLRRC aims to simultaneously cluster tasks and detect outlier tasks by integrating loss functions for tasks with robust regularization terms. The regularization term is inspired by the robust convex clustering \citep{Quan2020-mc}, which is further extended to incorporate non-convex and group-sparse penalties. This extension allows MTLRRC to effectively identify outlier tasks and achieve robust task clustering.
We further establish a connection between the extended robust clustering and the multivariate \textit{M}-estimator. This connection provides an intuitive interpretation of the robustness of MTLRRC against outlier tasks.
The parameters included in MTLRRC are estimated by a modified version of the alternating direction method of multipliers (ADMM; \cite{Boyd2011-ul}). 

The remainder of this paper is structured as follows. In Section \ref{Sec2},  we first illustrate the problem set-up of the MTL method.
After that, we review MTL methods based on convex clustering and introduce robust convex clustering along with its non-convex and group-sparse extensions. We also explore the connection between the extended robust clustering problem and the multivariate \textit{M}-estimator. In Section \ref{Sec3}, we present the MTLRRC and discuss its interpretation concerning the multivariate \textit{M}-estimator. In Section \ref{Sec4}, the modified ADMM algorithm for estimating the parameters of MTLRRC is established. Simulation studies are presented in Section \ref{Sec5} to demonstrate the effectiveness of MTLRRC and the application to real-world datasets is illustrated in Section \ref{Sec6}.  The conclusions are given in Section \ref{Sec7}.

\section{Motivation and methodology}
\label{Sec2}
\subsection{Problem set-up}
Suppose that we have $T$ datasets. For each dataset $m \ (m=1,\ldots,T)$, we observed  $n_{m}$ pairs of data points $\{(\bm{x}_{mi},y_{mi});i=1,\ldots,n_m\}$, where $\bm{x}_{mi}$ is a $p$-dimensional explanatory variables and $y_{mi}$ is the corresponding response variable following distribution in the exponential family with mean $\mu_{mi}=\mathbb{E}[y_{mi}\rvert \bm{x}_{mi}]$. Let $X_{m}=(\bm{x}_{m1},\ldots \bm{x}_{mn_{m}})^{\top}\in\mathbb{R}^{n_{m}\times p}$ be the design matrix and  $\bm{y}_{m}=(y_{m1},\ldots,y_{mn_{m}})^{\top}\in\mathbb{R}^{n_{m}}$ be the response vector for dataset $m$. We assume that each feature vector $\bm{x}_{mi}$ is standardized to have zero mean and unit variance and the response vector $\bm{y}_{m}$ is centered to have zero mean when it is a continuous.

Our goal is to estimate $T$ generalized linear models (GLMs) simultaneously, which take the form:
\begin{equation}
    \label{GLM}
  \eta_{mi}=g(\mu_{mi}) = w_{m0} + \bm{x}_{mi}^{\top}\bm{w}_{m}, \quad i=1,\ldots,n_{m},\;m=1,\ldots,T,
\end{equation}
where $w_{m0}$ is an intercept for $m$-th task, $\bm{w}_{m}=(w_{m1},\ldots,w_{mp})^\top$ is a $p$-dimensional regression coefficient vector for $m$-th task, $\eta_{mi}$ is a linear predictor, and $g(\cdot)$ is a canonical link function.
The density function of the response given the explanatory variables is expressed as 
\begin{equation}
    \nonumber
    f(y_{mi} \rvert \bm{x}_{mi};\theta(\bm{x}_{mi}),\phi) = \exp\left\{ \frac{y_{mi}\theta(\bm{x}_{mi})-b(\theta(\bm{x}_{mi}))}{a(\phi)}+c(y_{mi},\phi)\right\},
\end{equation}
where $a(\cdot),b(\cdot)$, and $c(\cdot)$ are known functions that vary according to the distributions, $\phi$ is a known dispersion parameter, and $\theta(\cdot)$ is the natural parameter, which is expressed as $g(\mu_{mi})=\theta(\bm{x}_{mi})$. 
In this paper, we assume that all $\bm{y}_m$ follow only the same type of distribution, regardless of the tasks.

Let $W=(\bm{w}_{1},\ldots,\bm{w}_{T})^{\top}\in\mathbb{R}^{T\times p}$ be the regression coefficient matrix, and $\bm{w}_{0}=(w_{10},\ldots,w_{T0})^{\top}\in\mathbb{R}^{T}$ be the intercept vector. 
To estimate $T$ GLMs in \eqref{GLM} simultaneously, we formulate the regularization-based multi-task learning (MTL) method as
\begin{equation}
        \label{MTL} 
        \min_{\substack{\bm{w}_{0}, W}}
                \left\{ \sum_{m=1}^{T} \frac{1}{n_{m}}L(w_{m0},\bm{w}_{m})+ \lambda \Omega(W) \right\}, 
\end{equation}
where $L(w_{m0},\bm{w}_{m})$ is a loss function derived from the negative log-likelihood of the $m$-th GLM task, $\lambda$ is a regularization parameter with a non-negative value, and $\Omega(\cdot)$ is a regularization term that encourages sharing the information among tasks. 
When continuous response vectors $\bm{y}_{m}\in\mathbb{R}^{n_{m}}$ are considered, the linear regression is given by $a(\phi)=\phi, \phi=1$, and $b(\theta) =\frac{\theta^{2}}{2}$. The regression loss function is
\begin{equation}
\nonumber
    L(w_{m0},\bm{w}_{m}) = \frac{1}{2}\|\bm{y}_{m} - \bm{X}_{m}\bm{w}_{m} \|_{2}^{2}.
\end{equation}
Note that the intercepts are excluded from the model without a loss of generality. 
When binary response vectors $\bm{y}_{m}\in\{0,1\}^{n_{m}}$ are considered, the logistic regression is given by $a(\phi)=\phi, \phi=1$, and $b(\theta) =\log(1+e^{\theta})$. The loss function of logistic regression is
\begin{equation}
\nonumber
    L(w_{m0},\bm{w}_{m}) = - \sum_{i=1}^{n_{m}}\left\{y_{mi}(w_{m0}+\bm{w}_{m}^{\top}\bm{x}_{i})-\log\{1+{\exp{(w_{m0}+\bm{w}_{m}^{\top}\bm{x}_{i})}}\} \right\}.
\end{equation}

\subsection{Multi-task learning via convex clustering}
To classify tasks into some clusters that share common characteristics,  
\cite{Okazaki2023-rc} proposed an MTL method called multi-task learning via convex clustering (MTLCVX) as follows:

\begin{equation}
\nonumber
    \begin{split}
        \label{MTLCVX} \min_{\bm{w}_{0}, W,U}
                \left\{ \sum_{m=1}^{T} \frac{1}{n_{m}}L(w_{m0},\bm{w}_{m})\right.&+\frac{\lambda_{1}}{2}\sum_{m=1}^{T}\|\bm{w}_{m}-\bm{u}_{m}\|_{2}^{2}\\
                &\left.+\lambda_{2}\sum_{(m_{1},m_{2})\in\mathcal{E}}r_{m_{1},m_{2}}\|\bm{u}_{m_{1}}-\bm{u}_{m_{2}} \|_{2} \right\}, \\
    \end{split}
\end{equation}
 where $\bm{u}_{m}=(u_{m1},\ldots,u_{mp})^{\top}\in\mathbb{R}^{p}$ is a centroid parameter for $m$-th task, $U=(\bm{u}_{1},\ldots,\bm{u}_{T})^{\top}$ is a $T\times p$ matrix, $r_{m_{1},m_{2}}$ is a weight between $m_{1}$-th and $m_{2}$-th task, $\mathcal{E}$ is a set of task pairs $(m_{1},m_{2})$, and $\lambda_{1}$ and $\lambda_{2}$ are non-negative regularization parameters. In this problem, the regularization terms are based on the convex clustering \citep{Hocking2011-mn}, which is to perform clustering of tasks. The second term encourages the centroids $\bm{u}_m$ to be close to their corresponding regression coefficients $\bm{w}_m$. The third term induces $\bm{u}_{m_{1}}\simeq\bm{u}_{m_{2}}$. If the value of $\bm{u}_{m_{1}}$ and $\bm{u}_{m_{2}}$ are estimated to be the same, $m_{1}$-th and $m_{2}$-th task are interpreted as belonging to the same cluster and sharing same characteristics. Therefore, $\bm{u}_{m}$ can be viewed as a center of cluster.

This minimization problem is solved by the block coordinate descent (BCD) algorithm. The algorithm is performed by alternately minimizing $\bm{w}_{m}$ and $\bm{u}_{m}$: we solve the regression model regularized by ridge penalty whose center is $\bm{u}_{m}$ and convex clustering where each $\bm{w}_{m} \;(m=1,\ldots, T)$ is considered as a sample, alternately. 
    
The weights $r_{m_{1},m_{2}}$ are set by the $k$-nearest neighbor as follows:
\begin{equation}
    \label{setR}
        R = \frac{S^{\top}+S}{2}, \quad (S)_{m_{1}m_{2}}=
        \begin{cases}
             1 & \widehat{\bm{w}}^{\mathrm{STL}}_{m_{1}}\;\mathrm{is}\; \mathrm{a}\;k\textrm{-nearest}\; \mathrm{neighbor} \; \mathrm{of}\;\widehat{\bm{w}}_{m_{2}}^{\mathrm{STL}}, \\
             0 & \mathrm{otherwise},
        \end{cases}
\end{equation} 
    where $R$ is a $\abs{\mathcal{E}}\times\abs{\mathcal{E}}$ matrix whose each component is $(R)_{m_{1}m_{2}}= r_{m_{1},m_{2}}$ and $\widehat{\bm{w}}^{\mathrm{STL}}_{m}$ is an estimated regression coefficient vector for $m$-th task by single-task learning such as  OLS, ridge, and lasso. This setting is also used in \cite{Yamada2017-cg}. This constructed weights would be based on that of convex clustering \citep{Hocking2011-mn}, which is used to reduce the computational costs by setting some $r_{m_{1},m_{2}}$ to zero and to improve the estimation results of clustering. 
    
    \subsection{Robust convex clustering}
    
    Suppose that we have $n$ observed $p$-dimensional data $\left\{\bm{x}_{i};i=1,\ldots,n\right\}$. To classify these data into distinct clusters, convex clustering (\cite{Pelckmans2005-kw}; \cite{Hocking2011-mn};  \cite{Lindsten2011-bs}) has been proposed. 
    
    \cite{Quan2020-mc} pointed out that convex clustering is sensitive to just a few outliers, and they proposed robust convex clustering (RCC) as follows:
     \begin{equation}
        \label{RCC}
        \min_{U,O}
        \left\{ \sum_{i=1}^{n} \frac{1}{2}\|\bm{x}_{i} - \bm{u}_{i} -\bm{o}_{i}\|_{2}^{2}+\lambda_{1}\sum_{(i_{1},i_{2})\in\mathcal{E}}r_{i_{1},i_{2}}\|\bm{u}_{i_{1}}-\bm{u}_{i_{2}} \|_{2} +\lambda_{2}\sum_{i=1}^{n}\|\bm{o}_{i}\|_{1} \right\},\\
    \end{equation}
  where $\bm{u}_{i}=(u_{i1},\ldots,u_{ip})^{\top}\in\mathbb{R}^{p}$ is a 
 vector of centroid parameters for $i$-th sample, $\bm{o}_{i}=(o_{i1},\ldots,o_{ip})^{\top}\in\mathbb{R}^{p}$ is a vector of outlier parameters for $i$-th sample, $U=(\bm{u}_{1},\ldots,\bm{u}_{n})^{\top}$ and $O=(\bm{o}_{1},\ldots,\bm{o}_{n})^{\top}$ are $n\times p$ matrices, respectively. The third term selects the outlier parameters by shrinking each element of $\bm{o}_{i}$ toward exactly zero. If $o_{ij}$ is estimated to be a non-zero value, the $j$-th feature of the $i$-th sample is considered an outlier. 
 
 From the relationship between the loss function of least square and $L_{1}$ penalty for the outlier parameters (\cite{Antoniadis2007-ir}; \cite{Gannaz2007-xf}),
  the minimization problem \eqref{RCC} is equivalent to the following minimization problem:
 \begin{equation}
 \label{RCC_huber}
      \min_{U}
        \left\{ \sum_{i=1}^{n}\sum_{j=1}^{p} h_{\lambda_{2}}(x_{ij} - u_{ij})+\lambda_{1}\sum_{(i_{1},i_{2})\in\mathcal{E}}r_{i_{1},i_{2}}\|\bm{u}_{i_{1}}-\bm{u}_{i_{2}} \|_{2} \right\},\\
 \end{equation}
 where $h_{\lambda}(\cdot)$ is a Huber's loss function defined as
 \begin{equation}
 \nonumber
     h_{\lambda_{2}}(z)=\begin{cases}
         \frac{1}{2}z^{2} & \abs{z} \leq \lambda,
         \\
       \lambda_{2} \abs{z}-\frac{\lambda^{2}}{2} & \abs{z}\geq \lambda.
      \end{cases}
 \end{equation}
 This would indicate that RCC is a robust version of convex clustering derived from replacing the loss function with the component-wise robust loss function.
 
\subsection{Non-convex extensions of robust convex clustering}
Although \cite{Quan2020-mc} only considered  $\ell_{1}$ penalty to select outlier parameters, it also can be considered other types of shrinkage penalties such as non-convex penalties and group penalties. For example, if group lasso \citep{Yuan2006-pd} is employed for the third term in \eqref{RCC}, we can detect the sample-wise outliers. Since our purpose in this paper is to detect task-wise outliers, we first consider the generalized robust clustering problem to detect sample-wise outliers as follows:
\begin{equation}
        \label{GRRC}
        \min_{U,O}
        \left\{ \sum_{i=1}^{n} \frac{1}{2}\|\bm{x}_{i} - \bm{u}_{i} -\bm{o}_{i}\|_{2}^{2}+\lambda_{1}\sum_{(i_{1},i_{2})\in\mathcal{E}}r_{i_{1},i_{2}}\|\bm{u}_{i_{1}}-\bm{u}_{i_{2}} \|_{2} + \sum_{i=1}^{n}P(\bm{o}_{i};\lambda_{2},\gamma) \right\},\\
\end{equation}
where $P(\cdot;\lambda,\gamma)$ is a penalty function that induces group sparsity and $\gamma$ is a tuning parameter that adapts the shape of the penalty function. By estimating $\bm{o}_{i}$ as a zero vector through group penalties, this problem aims to detect sample-wise outliers. For $i$-th sample $\bm{x}_{i}$, even if the value of one feature $x_{ij}$ has extensive value compared with its cluster center $\widehat{u}_{ij}$, $\widehat{\bm{o}}_{i}$ would not be non-zero vector. Only when the $L_{2}$-distance $\|||\bm{x}_{i}-\widehat{\bm{u}}_{i}\|_{2}$ has the extensive value, $\widehat{\bm{o}}_{i}$ is estimated to be non-zero vector and $\bm{x}_{i}$ is interpreted as an outlier sample.

If a non-convex penalty such as group SCAD and group MCP \citep{Huang2012-tk} is employed for the third term, the minimization problem is no longer a convex optimization problem. Therefore, we refer to the minimization problem as \textbf{R}obust \textbf{R}egularized \textbf{C}lustering (RRC). 

 \begin{algorithm}[H]
    \caption{Block coordinate descent algorithm for Problem \eqref{GRRC}}
    \label{BCD_GRRC}
    \begin{algorithmic}
      \Require $ X,\lambda_{1},\lambda_{2},\gamma,O^{(0)}$
      \While{until convergence of $U^{(t)}$ and $O^{(t)}$}
        \State 
        \begin{equation}
            \label{updata_U}
            U^{(t+1)} = \argmin_{U} L(U,O^{(t)})
        \end{equation}
        \State 
        \begin{equation}
            \label{updata_O}
            O^{(t+1)} = \argmin_{O} L(U^{(t+1)},O)
        \end{equation}    
       \EndWhile
      \Ensure $U,O$
    \end{algorithmic}
\end{algorithm}

The minimization problem \eqref{GRRC} is solved by the BCD algorithm shown in Algorithm \ref{BCD_GRRC}. Here, $X= (\bm{x}_{1},\ldots,\bm{x}_{n})^{\top}\in\mathbb{R}^{n \times p}$, $L(U, O)$ is the objective function in \eqref{GRRC}, and the superscripts in parentheses indicate the number of updates. The problem \eqref{updata_U} can be solved by algorithms for the convex clustering. 
Since the problem \eqref{updata_O} is separable in terms of $\bm{o}_{i}$, it is expressed as 
\begin{equation}
    \label{uni_group_opti}
  \bm{o}_{i}^{(t+1)} = \argmin_{\bm{o}_{i}} \left\{\frac{1}{2}\|\bm{x}_{i}-\bm{u}^{(t+1)}_{i}-\bm{o}_{i} \|_{2}^{2} + P(\bm{o}_{i};\lambda_{2},\gamma)\right\},\quad i=1,\ldots,n.
\end{equation}
Therefore, the update can be obtained by
\begin{equation}
\nonumber
    \bm{o}_{i}^{(t+1)} = \bm{\Theta}(\bm{x}_{i}-\bm{u}^{(t+1)}_{i};\lambda_{2},\gamma), \quad i=1,\ldots,n,
\end{equation}
where $\bm\Theta(\cdot;\lambda,\gamma)$ is a group-thresholding function defined for the corresponding penaly function $P(\cdot;\lambda,\gamma)$. For example, $\bm{\Theta}(\cdot;\lambda,\gamma)$ for group lasso is given by
\begin{equation}
\nonumber
    \bm{\Theta}^{\mathrm{glasso}}(\bm{z};\lambda,\gamma)=
    S(\bm{z};\lambda),
\end{equation}
where $S(\cdot;\lambda)$ is a group soft-thresholding function defined as 
\begin{equation}
\nonumber
S(\bm{z};\lambda) = \max\left(0,1-\frac{\lambda}{\|\bm{z}\|_{2}}\right)\bm{z}.
\end{equation}

The solution of \eqref{GRRC} obtained by Algorithm \ref{BCD_GRRC} is related to \textit{M}-estimators, which is similar to the connection between the minimization problems \eqref{RCC} and \eqref{RCC_huber}.
Let $A_{r}$ be a $\abs{\mathcal{E}}\times n$ matrix whose each row is $r_{i_{1},i_{2}}\bm{a}_{(i_{1},i_{2})}^{\top}$ and we set
\begin{equation}
\nonumber
    D = A_{r} \otimes I_{p},
\end{equation}
where $\otimes$ denotes the Kronecker product, $I_{p}$ is a $p\times p$ identity matrix, and $\bm{a}_{(i_{1},i_{2})}$ is defined as
    \begin{equation}
        \label{definition_of_A}
        (\bm{a}_{(i_{1},i_{2})})_{i}=
        \begin{cases}
           1 \quad   &i=i_{1},\\
           -1 \quad  &i=i_{2},\\
            0 \quad &\mathrm{otherwise},
        \end{cases}
         i=1,\ldots,n.
    \end{equation}
We also define the mixed $(2,1)$-norm \citep{Lounici2011-hm} for a $\abs{\mathcal{E}}p$-dimensional vector $\bm{z}$ as 
\begin{equation}
\nonumber
\|\bm{z} \|_{2,1} =  \sum_{k=1}^{\abs{\mathcal{E}}}\left(
\sum_{j= (k-1)p + 1}^{kp}z_{j}^{2}\right)^{1/2}.
\end{equation}
Using these definitions, the second term in \eqref{GRRC} can be written as
 \begin{equation}
 \nonumber
\sum_{(i_{1},i_{2})\in \mathcal{E}} r_{i_{1},i_{2}} \| \bm{u}_{i_{1}}-\bm{u}_{i_{2}} \|_{2} = \|D\mathrm{vec}(U) \|_{2,1}.
\end{equation}
Based on the above definitions, we summarize the relationship between the solution of the RRC and \textit{M}-estimator in the following proposition.

\begin{propo}
\label{propo1}
Suppose that $\widehat{U}$ is a convergence point in Algorithm \ref{BCD_GRRC} and $\bm{\psi}(\bm{o};\lambda,\gamma)=\bm{o} - \bm{\Theta}(\bm{o};\lambda,\gamma)$. Then, the $\widehat{U}$ satisfies the equation 
      \begin{equation}
      \nonumber
     -\Psi(X-\widehat{U};\lambda_{2},\gamma)+\lambda_{1}\frac{\partial}{\partial \mathrm{vec}(U)}(\|D\mathrm{vec}(U) \|_{2,1})\rvert_{U=\widehat{U}}= \bm{0},
 \end{equation}
  where $\Psi(X-\widehat{U};\lambda_{2},\gamma)$ is an $np$-dimensional vector defined as
 \begin{equation}
 \nonumber
     \Psi(X-\widehat{U};\lambda_{2},\gamma) = \begin{pmatrix} \bm{\psi} (\bm{x}_{1}-\bm{\widehat{u}}_{1};\lambda_{2},\gamma) \\ \bm{\psi} (\bm{x}_{2}-\bm{\widehat{u}}_{2};\lambda_{2},\gamma) \\ \vdots \\ \bm{\psi} (\bm{x}_{n}-\bm{\widehat{u}}_{n};\lambda_{2},\gamma) \end{pmatrix}.
 \end{equation}
\end{propo}
\noindent The proof is given in Appendix \ref{appendix:proofs1}. This proposition indicates that any local minimum point $\widehat{U}$ is one of the  stationary points of the following minimization problem:
\begin{equation}
    \label{GRRC_robustloss}
    \min_{U} \left\{\sum_{i=1}^{n} \rho_{\lambda_{2},\gamma}(\bm{x}_{i}-\bm{u}_{i})+ \lambda_{1}\sum_{(i_{1},i_{2})\in \mathcal{E}} r_{i_{1},i_{2}} \| \bm{u}_{i_{1}}-\bm{u}_{i_{2}} \|_{2}\right\},
\end{equation} 
where $\rho_{\lambda,\gamma}(\cdot)$ is a multivariate loss function that satisfies 
\begin{equation}
\nonumber
    \frac{\partial }{\partial \bm{z}} \rho_{\lambda,\gamma}(\bm{z}) = \bm{\psi}(\bm{z};\lambda,\gamma).
\end{equation} 
The sketch of some multivariate loss functions and corresponding group-thresholding functions is shown in Figure \ref{loss_function_thresholding_function}. Those formulations are summarized in Appendix \ref{app_loss}.

Proposition \ref{propo1} is inspired by similar propositions in \cite{She2011-gu} and \cite{Katayama2017-og}. While they considered linear regression case and only the situation where $\bm{\Theta}(\cdot;\lambda,\gamma)$ is defined as the component-wise thresholding function, we consider clustering problem and the situation where $\Theta(\cdot;\lambda,\gamma)$ is a group-thresholding function.  This enables us to solve clustering problems with multivariate robust loss functions by optimizing the problems with group penalties for outlier parameters instead. 

\begin{figure}[H]
    \begin{tabular}{ccc}
    \begin{minipage}[t]{0.33\hsize}
        \centering
        \includegraphics[width=4cm]{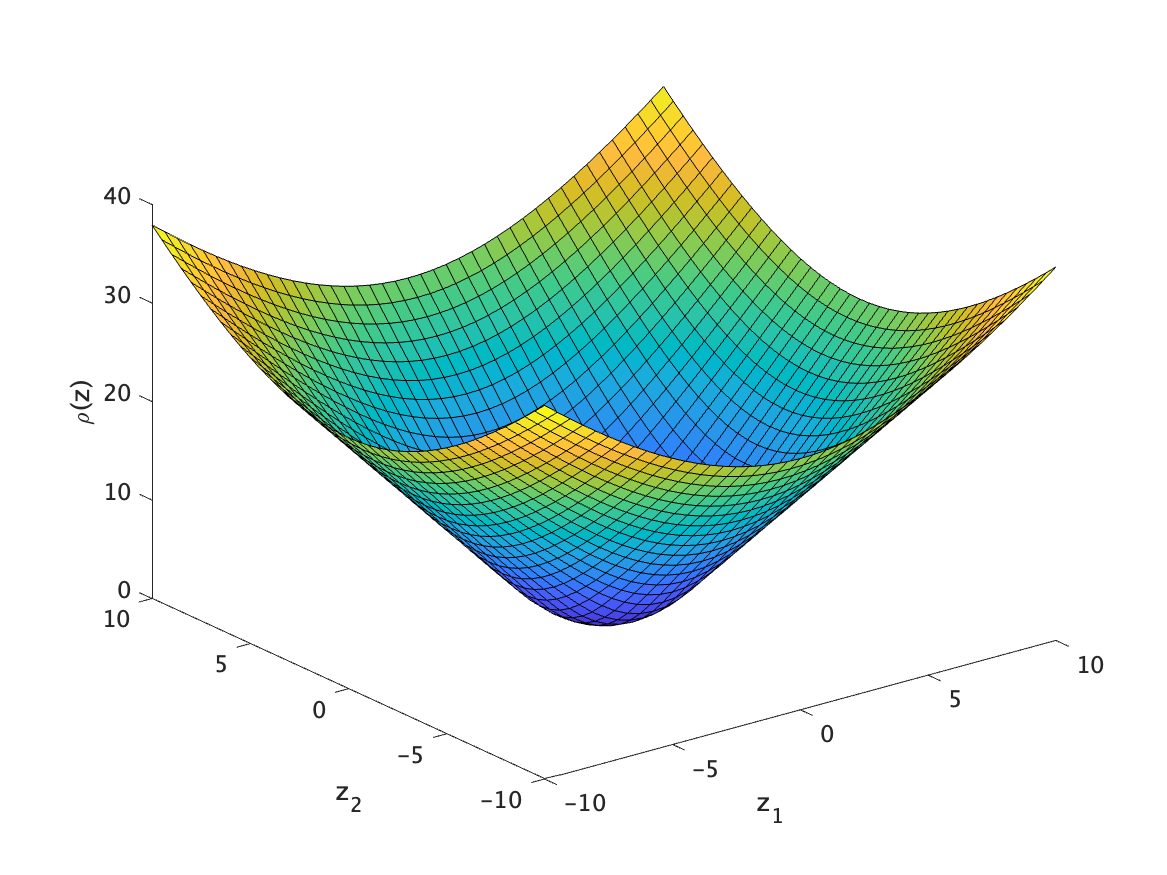}
        \subcaption{multivariate Huber's loss}
    \end{minipage}
    &
     \begin{minipage}[t]{0.33\hsize}
        \centering
        \includegraphics[width=4cm]{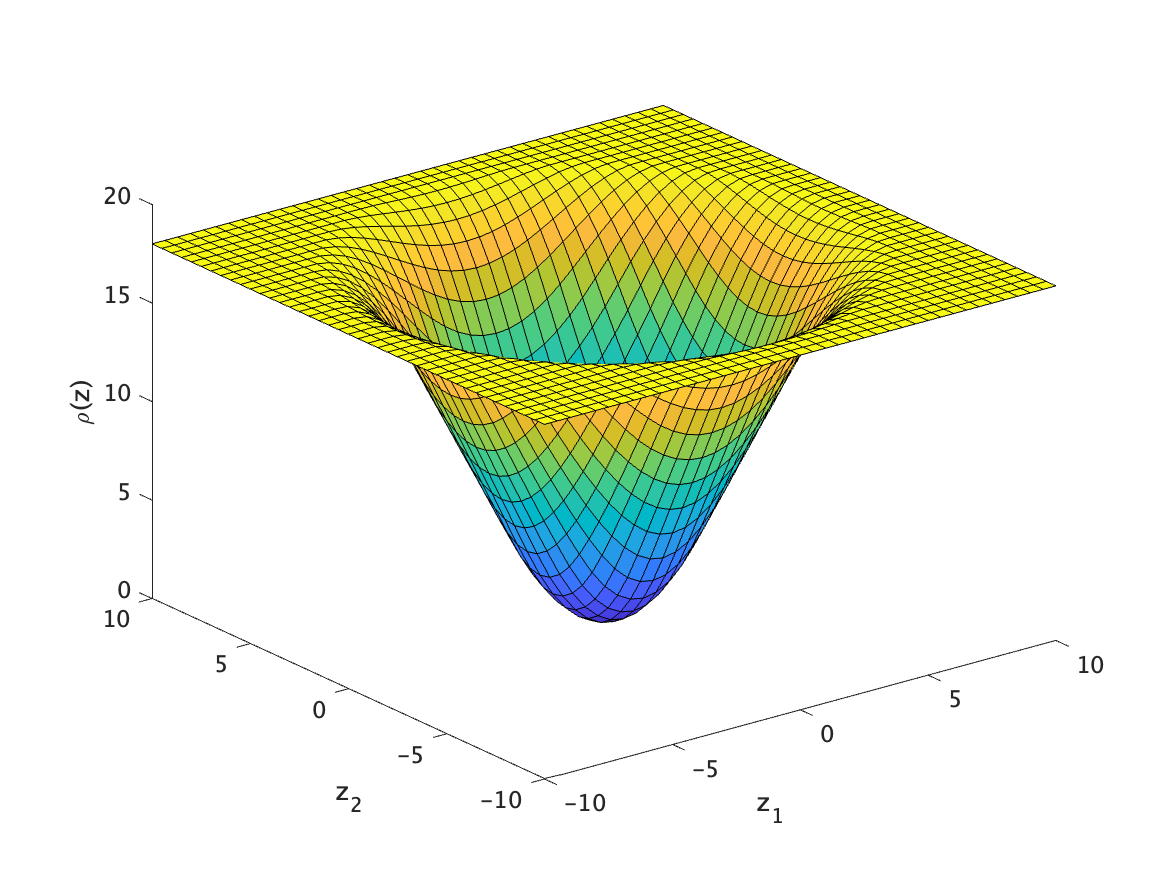}
        \subcaption{group SCAD loss}
      \end{minipage}
      &
    \begin{minipage}[t]{0.33\hsize}
        \centering
        \includegraphics[width=4cm]{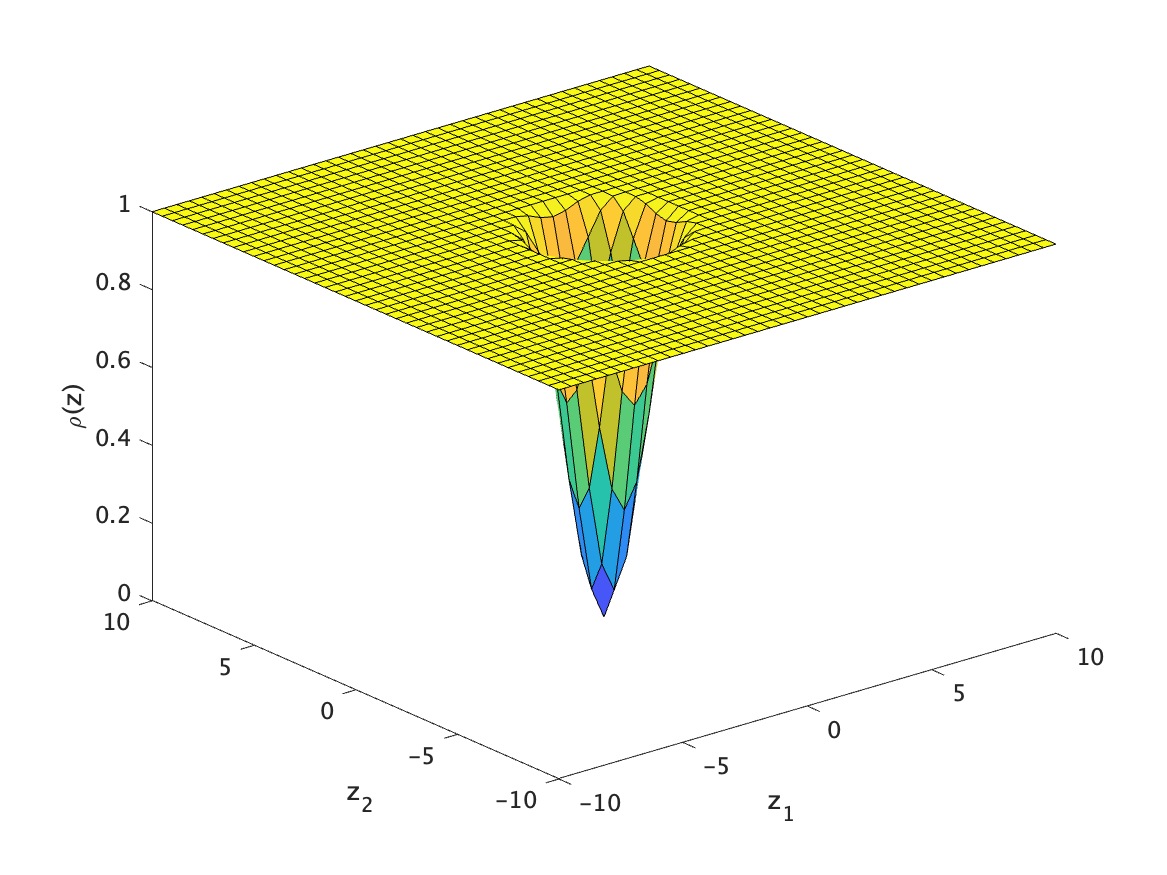}
        \subcaption{group Tukey's loss}
      \end{minipage}
    \\
    \begin{minipage}[t]{0.33\hsize}
        \centering
        \includegraphics[width=4cm]{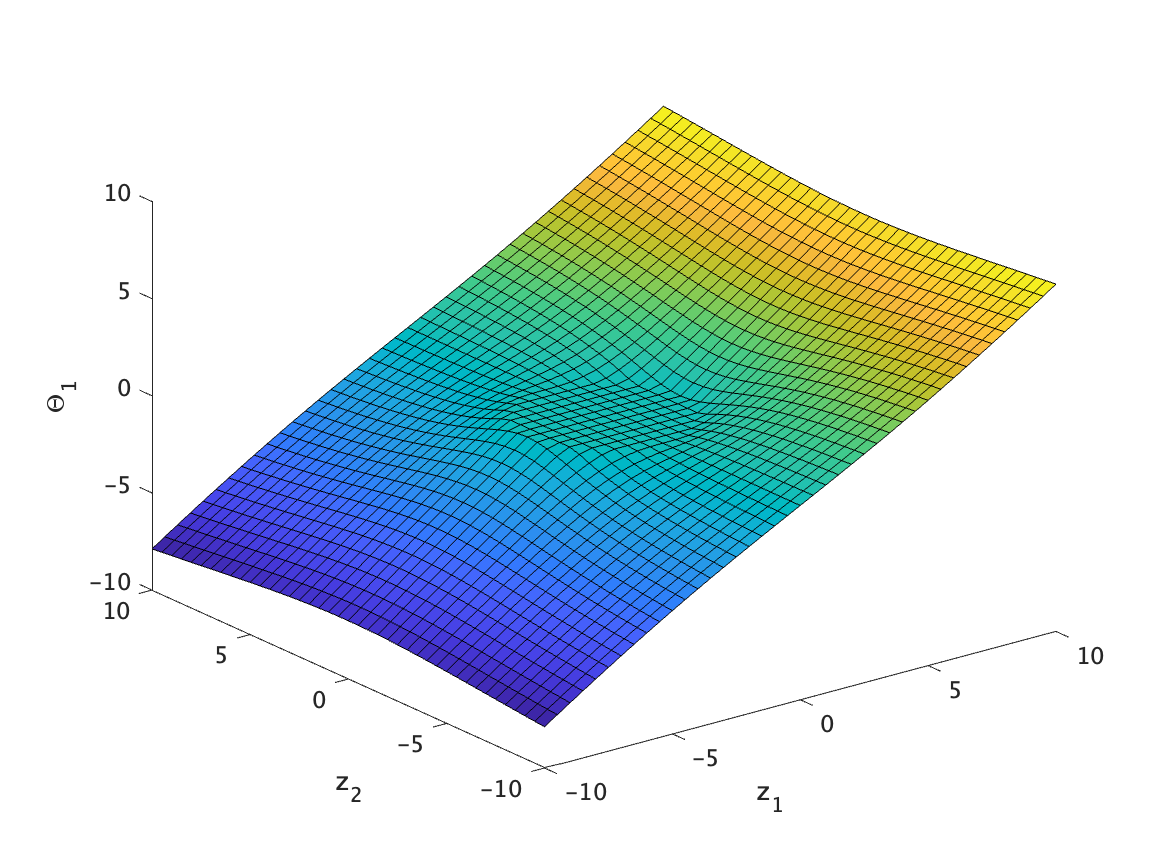}
        \subcaption{group soft thresholding}
      \end{minipage}
      &
          \begin{minipage}[t]{0.33\hsize}
        \centering
        \includegraphics[width=4cm]{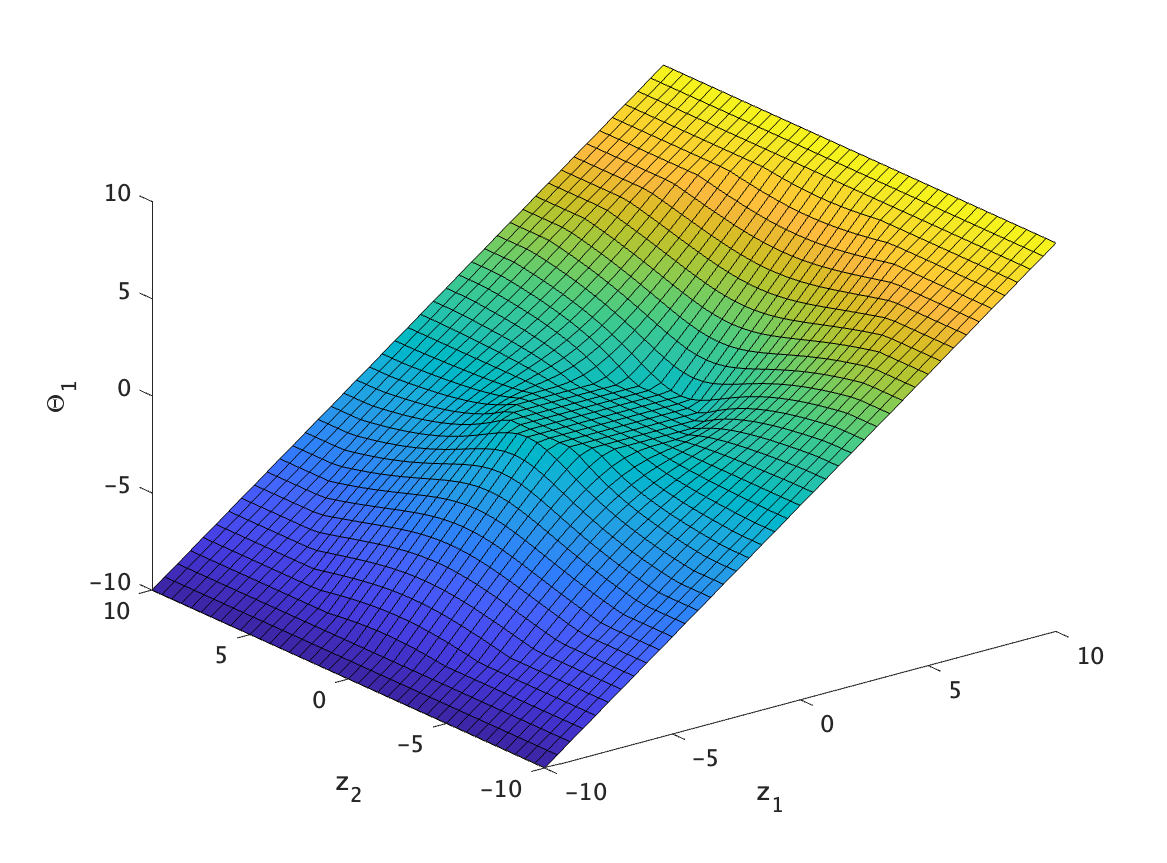}
        \subcaption{group SCAD thresholding}
      \end{minipage}
      &
          \begin{minipage}[t]{0.33\hsize}
        \centering
        \includegraphics[width=4cm]{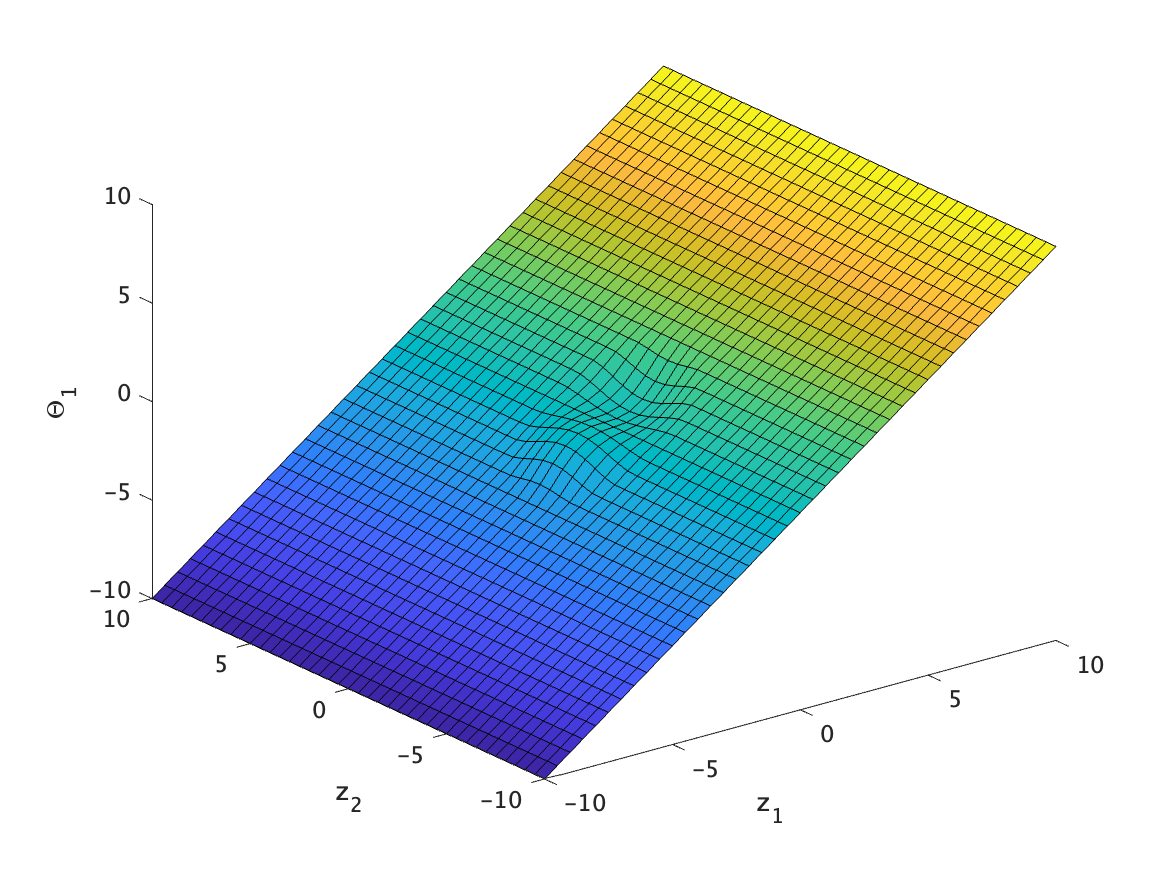}
        \subcaption{group Tukey's thresholding}
      \end{minipage}
    \end{tabular}
    \caption{The multivariate loss functions (top row), the group-thresholding functions (bottom row). The x-axis and y-axis represent the values of input $\bm{z}\in\mathbb{R}^{2}$. The z-axis shows the first component of output $\Theta(\bm{z};\lambda,\gamma)$ in the top row, and the $\rho_{\lambda,\gamma}(\bm{z})$ in the bottom row. The values of $\lambda$ and $\gamma$ are fixed with three.}
    \label{loss_function_thresholding_function}
\end{figure}

\section{Proposed method}
\label{Sec3}
\subsection{Multi-task learning via robust regularized clustering}

Existing MTL methods based on clustering have not taken into account the presence of outlier tasks. For instance, MTLCVX estimates the regression coefficients uniformly close to the corresponding estimated centroids, which ignores the nature of outlier tasks that have a large task-specific characteristic or no relationship to other tasks. 
Some robust MTL methods \citep{Chen2011-yr, Gong2012-gl} have attempted to address the issues of outlier tasks by introducing outlier parameters and selecting them using group lasso regularization. However, group lasso \citep{Yuan2006-pd} limits the value of the outlier parameters, which may not adequately represent their nature.

To overcome these problems, we propose the \textbf{M}ulti-\textbf{T}ask \textbf{L}earning via \textbf{R}obust \textbf{R}egularized \textbf{C}lustering (MTLRRC).
MTLRRC is formulated as follows:
\begin{equation}
    \label{MTLRRC}
        \begin{split}
        \min_{W,U,O}
         & \left\{ \sum_{m=1}^{T}\frac{1}{n_{m}}L(w_{m0},\bm{w}_{m}) +\frac{\lambda_{1}}{2}\sum_{m=1}^{T}\|\bm{w}_{m}-\bm{u}_{m}-\bm{o}_{m}\|_{2}^{2} \right. \\
        & \left. +\lambda_{2}\sum_{(m_{1},m_{2})\in\mathcal{E}}r_{m_{1},m_{2}}\|\bm{u}_{m_{1}}-\bm{u}_{m_{2}} \|_{2} + \sum_{m=1}^{T} P( \bm{o}_{m};\lambda_{3},\gamma) \right\},
        \end{split}
\end{equation}
where $\bm{o}_{m}=(o_{m1},\ldots,o_{mp})^{\top}\in\mathbb{R}^{p}$ is a vector of outlier parameters for $m$-th task, $\lambda_{3}$ is a regularization parameter with a non-negative value. The second through
the fourth term is based on the minimization problem \eqref{GRRC}. If $\bm{o}_{m}$ is estimated to be a non-zero vector, $m$-th task is considered to be an outlier task that does not share a common structure with any tasks. We set the weights $r_{m_{1},m_{2}}$ based on Eq. \eqref{setR}.

\subsection{Interpretation through the BCD algorithm}
\subsubsection{Convex case}
First, we consider that group lasso is employed for $P(\cdot;\lambda,\gamma)$ in \eqref{MTLRRC}. 
Then, since MTLRRC is a convex optimization problem, we can obtain another representation for \eqref{MTLRRC} by minimizing in terms of $O$ as follows:
\begin{equation}
    \label{MTLRRC_huber}
    \begin{split}
        \min_{\bm{w}_{0},W,U }\left\{
        \sum_{m=1}^{T} \frac{1}{n_{m}}L(w_{m0},\bm{w}_{m}) \right. &+\lambda_{1}\sum_{m=1}^{T}h_{\lambda_{3}/\lambda_{1}}^{\mathrm{M}}(\bm{w}_{m}-\bm{u}_{m})  \\
         & \left. +\lambda_{2}\sum_{(m_{1},m_{2})\in\mathcal{E}}r_{m_{1},m_{2}}\|\bm{u}_{m_{1}}-\bm{u}_{m_{2}} \|_{2} \right\},
    \end{split}
\end{equation}
where $h_{\lambda}^{\mathrm{M}}(\cdot)$ is a multivariate Huber's loss function \citep{Hampel1986-pa} defined as
\begin{equation}
\nonumber
 h_{\lambda}^{\mathrm{M}}(\bm{z})=
 \begin{cases}
    \frac{1}{2}\|\bm{z}\|_{2}^{2} & \|\bm{z}\|_{2}\leq \lambda,\\
    \lambda\|\bm{z}\|_{2}-\frac{\lambda^{2}}{2} & \|\bm{z}\|_{2}> \lambda.
 \end{cases}
\end{equation}
This representation helps us understand the interpretation of the proposed method \eqref{MTLRRC}. 

 \begin{algorithm}[htbp]
    \caption{Block coordinate descent algorithm for Problem \eqref{MTLRRC_huber}}
    \label{BCD_MTLRRC_huber}
    \begin{algorithmic}
      \Require $ (\bm{y}_{m},X_{m};m=1,\ldots,T),R,\lambda_{1},\lambda_{2},\lambda_{3},U^{(0)}$
      \While{until convergence of $W^{(t)}$ and $U^{(t)}$}
        \State 
        \begin{equation}
        \label{updata_W_BCD2}
            (\bm{w}_{0}^{(t+1)},W^{(t+1)}) = \argmin_{\bm{w}_{0},W} L^{\mathrm{MH}}(\bm{w}_{0},W,U^{(t)})
        \end{equation}
        \State 
        \begin{equation}
            \label{updata_U_BCD2}
            U^{(t+1)} = \argmin_{U} L^{\mathrm{MH}}(\bm{w}_{0}^{(t+1)},W^{(t+1)},U)
        \end{equation}    
       \EndWhile
      \Ensure $W,U$
    \end{algorithmic}
\end{algorithm}

Let the objective function of the minimization problem \eqref{MTLRRC_huber} be $L^{\mathrm{MH}}(\bm{w}_{0},W,U)$. We consider solving the minimization problem \eqref{MTLRRC_huber} by Algorithm \ref{BCD_MTLRRC_huber} based on the BCD algorithm. Since the minimization in terms of $\bm{w}_{m}$ is separable, the update \eqref{updata_W_BCD2} is expressed as 
\begin{equation}
\nonumber
    (w_{0}^{(t+1)},\bm{w}_{m}^{(t+1)}) = \argmin_{w_{0},\bm{w}_{m}}\left\{ \frac{1}{n_{m}}L(w_{m0},\bm{w}_{m}) + \lambda_{1} h_{\lambda_{3}/\lambda_{1}}^{\mathrm{M}}(\bm{w}_{m}-\bm{u}_{m}^{(t)}) \right\}, \quad m=1,\ldots,T.
\end{equation}
These updates estimate the regression coefficients for the $m$-th task to be close to the corresponding centroid. However, when $\|\bm{w}-\bm{u}_{m}^{(t)}\|_{2}$ tends to take a larger value than $\lambda_{3}/\lambda_{1}$, the shrinkage toward the centroid is reduced by the part of $\ell_{2}$-norm in the multivariate Huber's function. Thus, if $m$-th task is an outlier task, the estimated $\widehat{\bm{w}}_{m}$ is expected to be less affected by the common structure $\widehat{\bm{u}}_{m}$.

 The minimization problem in terms of the update \eqref{updata_U_BCD2} is in the framework of the minimization problem \eqref{GRRC_robustloss}. This can be seen by replacing $(\bm{x}_{i};i=1,\ldots,n)$ in \eqref{GRRC_robustloss} with $(\bm{w}_{m}^{(t+1)};m=1,\ldots,T)$ and $\rho_{\lambda_{2},\gamma}(\cdot)$ with $h_{\lambda_{3}}^{\mathrm{HM}}(\cdot)$. Consequently, the update of $U^{(t+1)}$ is performed under the robust clustering of tasks. Based on the discussions of Algorithm \ref{BCD_MTLRRC_huber}, the estimated values $\widehat{W}$ and $\widehat{U}$ in MTLRRC can be regarded as a convergence point of alternative estimation, which consists of a regression step that reduces shrinkage of outlier tasks toward cluster center and a robust clustering step for tasks. Therefore, MTLRRC is expected to be robust to the outlier tasks.

\subsubsection{Non-convex case}

Next, we consider that non-convex group penalties are employed for $P(\cdot;\lambda,\gamma)$ in \eqref{MTLRRC}. 
The estimates of the parameters can be calculated by Algorithm \ref{BCD_MTLRRC}. 
Here, $L^{\mathrm{MR}}(\bm{w}_{0},W,U,O)$ is the objective function of the minimization problem \eqref{MTLRRC}. 
Then, the following proposition similar to Proposition \ref{propo1} holds. 
\begin{propo}
\label{propo2} Let $\bm{w}_{m}^{\prime}=(w_{m0},\bm{w}_{m}^{\top})^{\top}$.
Suppose that $\left(\widehat{\bm{w}}_{0},\widehat{W},\widehat{U}\right)$ is a pair of convergence point in Algorithm \ref{BCD_MTLRRC} and $\bm{\psi}(\bm{o};\lambda,\gamma)=\bm{o} - \bm{\Theta}(\bm{o};\lambda,\gamma)$. Then, $(\widehat{\bm{w}}_{0},\widehat{W},\widehat{U})$ satisfies  
\begin{align}
\label{propo2_regression}
\frac{\partial}{\partial \bm{w}_{m}^{\prime}}\frac{1}{n_{m}}L(w_{m0},\bm{w}_{m})\rvert_{\bm{w}_{m}^{\prime}=\widehat{\bm{w}}_{m}^{\prime}}+ \lambda_{1}
\begin{pmatrix}
    0 \\
\bm{\psi}(\widehat{\bm{w}}_{m}-\widehat{\bm{u}}_{m};\lambda_{3}/\lambda_{1},\gamma)    
\end{pmatrix}
 &= \bm{0},\quad m=1,\ldots,T, \\
\label{propo2_clustering}
     -\lambda_{1}\Psi(\widehat{W}-\widehat{U};\lambda_{3}/\lambda_{1},\gamma)+\lambda_{2}\frac{\partial}{\partial \mathrm{vec}(U)}(\|D\mathrm{vec}(U) \|_{2,1})\rvert_{U=\widehat{U}} &= \bm{0}.
\end{align}
\end{propo}
\noindent The proof is given in Appendix \ref{appendix:proofs2}. The equations \eqref{propo2_regression} and \eqref{propo2_clustering} are the same first-order conditions for the following minimization problem:
\begin{equation}
\nonumber
    \begin{split}
        \min_{\substack{\bm{w}_{m},\bm{u}_{m} \in \mathbb{R}^{p},\\m=1,\ldots,T}}\left\{
        \sum_{m=1}^{T} \frac{1}{n_{m}}L(\bm{w}_{m},w_{m0}) \right. &+\lambda_{1}\sum_{m=1}^{T}\rho_{\lambda_{3}/\lambda_{1},\gamma}(\bm{w}_{m}-\bm{u}_{m})  \\
         & \left. +\lambda_{2}\sum_{(m_{1},m_{2})\in\mathcal{E}}r_{m_{1},m_{2}}\|\bm{u}_{m_{1}}-\bm{u}_{m_{2}} \|_{2} \right\}.
    \end{split}
\end{equation}
Therefore, we may expect that the solution in the case of non-convex penalties has a similar interpretation of the minimization problem \eqref{MTLRRC_huber}.

\begin{algorithm}[t]
    \caption{Block coordinate descent algorithm for MTLRRC} 
    \label{BCD_MTLRRC}
    \begin{algorithmic}
      \Require $ (\bm{y}_{m},X_{m};m=1,\ldots,T),R,\lambda_{1},\lambda_{2},\lambda_{3},U^{(0)},O^{(0)}$
      \While{until convergence of $\bm{w}_{0}^{(t)},W^{(t)}$, $U^{(t)}$ and $O^{(t)}$}
        \State 
        \begin{equation}
        \nonumber
        (\bm{w}_{0}^{(t+1)},W^{(t+1)}) = \argmin_{\bm{w}_{0},W} L^{\mathrm{MR}}(\bm{w}_{0},W,U^{(t)},O^{(t)})
        \end{equation}
        \State 
        \begin{equation}
                \nonumber
            U^{(t+1)} = \argmin_{U} L^{\mathrm{MR}}(\bm{w}_{0}^{(t+1)},W^{(t+1)},U,O^{(t)})
        \end{equation}    
        \begin{equation}
                \nonumber
            O^{(t+1)} = \argmin_{O} L^{\mathrm{MR}}(\bm{w}_{0}^{(t+1)},W^{(t+1)},U^{(t+1)},O)
        \end{equation}
       \EndWhile
      \Ensure $\bm{w}_{0},W,U,O$
    \end{algorithmic}
\end{algorithm}

\section{Estimation algorithm via modified ADMM}
\label{Sec4}

  MTLRRC can be estimated by Algorithm \ref{BCD_MTLRRC}. However, this estimation procedure is computationally expensive, because the update of $U^{(t+1)}$ involves solving the convex clustering, which is computationally demanding. To avoid this computation, we consider estimating parameters included in MTLRRC by alternating direction method of multipliers (ADMM; \cite{Boyd2011-ul}).
  
  We consider the following minimization problem equivalent to Problem \eqref{MTLRRC}:
\begin{equation}
\nonumber
        \begin{split}
        \min_{\bm{w}_{0},W,U,O}
         & \left\{ \frac{1}{n_{m}}L(\bm{w}_{m0},\bm{w}_{m}) +\frac{\lambda_{1}}{2}\sum_{m=1}^{T}\|W-U-O\|_{F}^{2} \right. \\
        & \left. +\lambda_{2}\sum_{(m_{1},m_{2})\in\mathcal{E}}r_{m_{1},m_{2}}\|\bm{b}_{(m_{1},m_{2})} \|_{2} + \sum_{m=1}^{T} P( \bm{o}_{m};\lambda_{3},\gamma) \right\}, \\
        &\mathrm{s.t.} \quad \bm{u}_{m_{1}}-\bm{u}_{m_{2}}= \bm{b}_{(m_{1},m_{2})} ,\quad (m_{1},m_{2})\in\mathcal{E}.
        \end{split}
\end{equation}
For this minimization problem, we consider the following augmented Lagrangian:
\begin{equation}
    \label{augmented_lagrangian_MTLRRC}
    \begin{split}
        L_{\nu} (\bm{w}_{0},W,U,O,B,S) & = \sum_{m=1}^{T}\frac{1}{n_{m}}L(w_{m0},\bm{w}_{m}) +\frac{\lambda_{1}}{2}\|W-U-O\|_{F}^{2}  \\
        &+\lambda_{2}\sum_{(m_{1},m_{2})\in\mathcal{E}}r_{m_{1},m_{2}}\|\bm{b}_{(m_{1},m_{2})} \|_{2} + \sum_{m=1}^{T} P( \bm{o}_{m};\lambda_{3},\gamma) \\
        &+ \mathrm{tr}(S^{\top}(B-A_{\mathcal{E}}U)) + \frac{\nu}{2} \|B-A_{\mathcal{E}}U\|_{F}^{2},
    \end{split}
\end{equation}
where $\mathrm{tr}(\cdot)$ denotes the trace operator, $A_{\mathcal{E}}$ is a $\abs{\mathcal{E}}\times T$ matrix whose each row is $\bm{a}_{m_{1},m_{2}}^{\top}$ defined with the same manner of Eq. \eqref{definition_of_A}, $S$ is a $\abs{\mathcal{E}}\times p$ Lagrangian multipliers matrix, and $\nu$ is a tuning parameter with non-negative value.

For this augmented Lagrangian, we consider the following updates of the modified ADMM:
\begin{equation}
\nonumber
    \begin{split}
        (\bm{w}_{0}^{(t+1)},W^{(t+1)}) &= \argmin_{\bm{w}_{0},W} L_{\nu} (\bm{w}_{0},W,U^{(t)},O^{(t)},B,S^{(t)}), \\
        U^{(t+1)} &= \argmin_{U} \left(\min_{B}L_{\nu}(\bm{w}_{0}^{(t+1)},W^{(t+1)},U,O^{(t)},B,S^{(t)})\right), \\
        O^{(t+1)} &= \argmin_{O} L_{\nu} (\bm{w}_{0}^{(t+1)},W^{(t+1)},U^{(t+1)},O^{(t)},B,S^{(t)}), \\
        S^{(t+1)}_{(m_{1},m_{2})} &= \mathrm{prox}((S^{(t)}+\nu A_{\mathcal{E}} U^{(t+1)})_{(m_{1},m_{2})},\lambda_{2}r_{m_{1},m_{2}}),\quad (m_{1},m_{2})\in\mathcal{E}, \\
     \end{split}
\end{equation}
where $\mathrm{prox}(\cdot,\lambda)$ is defined as
\begin{equation}
    \label{prox_group}
     \mathrm{prox}(\bm{z},\lambda)=\mathrm{min}(\|\bm{z}\|_{2},\lambda)\frac{\bm{z}}{\|\bm{z}\|_{2}}.
\end{equation}

The minimization problem in terms of $U$ and $B$ is jointly done. The minimization in terms of $B$ can be written explicitly, and we only need to solve that in terms of $U$ using the gradient method. These modifications about updates of $U, B$, and $S$ are based on the idea of \cite{Shimmura2022-ar}. Those details and derivation are provided in Appendix \ref{appendix_ADMM}. 

 The update for $\bm{w}_{0}$ and $W$ is given by solving the independent regularized GLMs for each task, which are expressed as 
\begin{equation}
\nonumber
    (\bm{w}_{m0}^{(t+1)},\bm{w}_{m}^{(t+1)}) = \argmin_{\bm{w}_{m0},\bm{w}_{m}}\left\{ \frac{1}{n_{m}}L(w_{m0},\bm{w}_{m}) + \frac{\lambda_{1}}{2}\|\bm{w}_{m}-\bm{u}_{m}^{(t)}-\bm{o}_{m}^{(t)} \|_{2}^{2} \right\},\;m=1,\ldots,T.
\end{equation}
These minimization problems are solved by the Newton-Raphson method provided in Algorithm \ref{Newton-Raphson}. 
The update of $O$ is given by the same manner as \eqref{uni_group_opti}. 
As a result, we obtain the estimation algorithm for MTLRRC as Algorithm \ref{MTLRRC_ADMM}. Here, $\mathrm{STL}(\cdot,\cdot)$ is a function returning an estimated regression coefficient vector by an arbitrary single-task learning method.

 Because the minimization problem \eqref{MTLRRC} includes the non-separable term concerning $U, W$, and $O$ for the second term and the non-convex term for the fourth term, the convergence of the ADMM algorithm is not guaranteed theoretically.
 However, we found that Algorithm \ref{MTLRRC_ADMM} converges to a point empirically in almost all simulation studies and applications to real data shown later.
  \begin{algorithm}[htbp]
    \caption{Newton-Raphson method for updating $w_{0}$ and $\bm{w}_{m}$}
    \label{Newton-Raphson}
    \begin{algorithmic}
      \Function {NR}{$n,X,\bm{y},\bm{u},\bm{o},\lambda_{1}$}
            \State Initialize; Let $\bm{X}^{\prime}=(\bm{1},X), \bm{w}^{\prime}=(w_{0},\bm{w}^{\top})^{\top}, \bm{u}^{\prime}=(0,\bm{u}^{\top})^{\top}$,
            \State $\bm{o}^{\prime}=(0,\bm{o}^{\top})^{\top},\Lambda= \mathrm{diag}(0,\lambda_{1},\ldots,\lambda_{1}),$
            \State $\bm{\mu} =\left(\frac{\partial b(\eta_{1})}{\partial \eta_{1}},\ldots,\frac{\partial b(\eta_{n})}{\partial \eta_{n}}\right)^{\top}, E =\mathrm{diag}\left(\frac{\partial^{2}b(\eta_{1})}{\partial\eta_{1}^{2}},\ldots,\frac{\partial^{2}b(\eta_{n})}{\partial\eta_{n}^{2}}\right)$.
      \While{until convergence of $\bm{w}^{\prime}$}
        \State $\bm{w}^{\prime(t+1)} = \bm{w}^{\prime(t)} + \left(\frac{X^{\prime\top}E^{(t)} X^{\prime}}{na(\phi)} +\Lambda\right)^{-1}\left\{\frac{X^{\prime\top}(\bm{y}-\bm{\mu}^{(t) })}{n a(\phi)}-\Lambda(\bm{w}^{\prime(t)}-\bm{u}^{\prime}-\bm{o}^{\prime})\right\}$
      \EndWhile
      \State Output: $(w_{0},\bm{w}^{\top})^{\top}=\bm{w}^{\prime}$
    \EndFunction
    \end{algorithmic}
  \end{algorithm}
    
    \begin{algorithm}[htbp]
    \caption{Estimation algorithm of MTLRRC via modified ADMM}
    \label{MTLRRC_ADMM}
    \begin{algorithmic}
      \Require $\{\bm{y}_{m},X_{m}; m=1,\ldots,T\},k,\lambda_{1},\lambda_{2},\lambda_{3},\gamma, U^{(0)},O^{(0)}$
      \For{ $m=1,\ldots,T$}
        \State $\widehat{\bm{w}}_{m}^{\mathrm{STL}}=  \mathrm{STL}(y_{m},X_{m})$
      \EndFor
      \State calculating $R$ by Eq. (\ref{setR}) from $k$ and $\widehat{\bm{w}}_{m}^{\mathrm{STL}}$
      \State converting $R$ into $A_{\mathcal{E}}$ by Eq. (\ref{definition_of_A})
      \State $G=A_{\mathcal{E}}^{\top}A_{\mathcal{E}},\iota=\frac{1}{\lambda_{1}+2\max_{i=1,\ldots,T}((G)_{ii})}$    
  
       \While{until convergence of $W^{(t)}$}
       \State update of $\bm{w}_{0}$ and $W$
         \For{ $m=1,\ldots,T$} 
             \State $(w_{m0}^{(t+1)},\bm{w}_{m}^{(t+1)\top})^{\top}  = \mathrm{NR}(n_{m},X_{m},\bm{y}_{m},\bm{u}_{m}^{(t)},\bm{o}_{m}^{(t)},\lambda_{1})$
         \EndFor
      \State update of $U$
      \State $l=0,\alpha^{(0)} =1, H^{(0)}=U^{(t)},C^{(0)}=U^{(t)}$
      \While{until convergence of $H^{(l)}$}
        \For{ $(m_{1},m_{2})\in\mathcal{E}$}
            \State $ F_{(m_{1},m_{2})}=\mathrm{prox}((S^{(t)}+ \nu A_{\mathcal{E}} \cdot C^{(l)})_{m_{1},m_{2}},\lambda_{2}r_{m_{1},m_{2}})$
         \EndFor
        \State $H^{(l+1)} = C^{(l)}-\iota\{\lambda_{1}(C^{(l)}+O^{(t)}-W^{(t+1)})+A_{\mathcal{E}}^{\top}F\} $
        \State $\alpha^{(l+1)}=\frac{1+\sqrt{1+4(\alpha^{(l)})^{2}}}{2}$
        \State $C^{(l+1)} =C^{(l)}+\frac{\alpha^{(l)}-1}{\alpha^{(l+1)}}(H^{(l+1)}-H^{(l)})$
      \EndWhile
       \State $U^{(t+1)}=H^{(l)}$
       \State update of $O$
           \For{$m=1,\ldots,T$}
            \State $\bm{o}_{m}^{(t+1)}= \bm{\Theta}(\bm{w}_{m}^{(t+1)}-\bm{u}_{m}^{(t+1)};\lambda_{3}/\lambda_{1},\gamma)$
         \EndFor
        \State update of $S$
         \For{$(m_{1},m_{2})\in\mathcal{E}$}
            \State $S^{(t+1)}_{(m_{1},m_{2})} = \mathrm{prox}((S^{(t)}+\nu A_{\mathcal{E}} \cdot U^{(t+1)})_{(m_{1},m_{2})},\lambda_{2}r_{m_{1},m_{2}})$
         \EndFor
      \EndWhile
      \Ensure $W,U,O$
    \end{algorithmic}
  \end{algorithm}

\section{Simulation studies}
\label{Sec5}

In this section, we report simulation studies in the linear regression setting. We generated data by the true model:
    \begin{equation}
    \nonumber
        \begin{split}
            \bm{y}_{m} = X_{m}\bm{w}_{m}^{\ast}+\bm{\epsilon}_{m}, \quad m=1,\ldots,T,
        \end{split}
    \end{equation}
where $\bm{\epsilon}_{m}$ is an error term whose each component is distributed as $N(0,\sigma^{2})$ independently, $X_{m}$ is a design matrix generated from $N_{p}(\bm{0},I_{p})$ independently, and $\bm{w}^{\ast}_{m}$ is a true regression coefficient vector for $m$-th task. 
For this true model, $T$ tasks consist of $C$ true clusters and other outlier tasks. First, all tasks were assigned to $C$ clusters with the same number of tasks in each cluster as $T/C$. Then, some of them were randomly assigned to outlier tasks. 

For the true structure of regression coefficient vectors,
we considered the following two cases:
\begin{equation}
\nonumber
\begin{split}
\mbox{Case 1:}&\quad
\bm{w}_{m}^{\ast}=\bm{u}_{c}^{\ast} + \bm{v}_{m}^{c\ast}+I(\tau_{m}=1)\bm{o}_{m}^{c\ast}, \\
 \mbox{Case 2:}&\quad
    \bm{w}_{m}^{\ast} = \begin{cases}
       \bm{u}_{c}^{\ast} + \bm{v}_{m}^{c\ast} & \mathrm{if}\;\tau_{m} =0, \\
       \bm{o}_{m}^{\ast} & \mathrm{if}\; \tau_{m} = 1,
    \end{cases}
    \end{split}
\end{equation}
where $\bm{u}_{c}^{\ast}$ is a true cluster center for $c$-th cluster, $\bm{v}_{m}^{c{\ast}}$ is a true task-specific parameter for $m$-th task belonging to $c$-th cluster, $\bm{o}_{m}$ is a true outlier parameter for $m$-th task, and $\tau_{m}$ is a random variable distributed as $ P(\tau_{m}=1)=\kappa$ and $P(\tau_{m}=0)=1-\kappa$.
$\tau_{m}=1$ means that the $m$-th task is assigned to an outlier task.
 Case 1 considers a situation where outlier tasks share the same cluster center with other tasks but the outlier parameter is added.
On the other hand, Case 2 considers a situation where outlier tasks do not have any common structure with other tasks.

The parameters $\bm{u}_{c}^{\ast},\bm{v}_{m}^{c\ast}$, and $\bm{o}_{m}^{\ast}$ were generated as follows. 
 First, each explanatory variable $\{j=1,\ldots,p\}$ was randomly assigned to the $c$-th clusters $\{c=1,\ldots,C \}$ with the same probability. Then, we generated a true centroid parameter for $c$-th cluster $\bm{u}^{\ast}_{c} = (u^{\ast}_{c1},\ldots,u^{\ast}_{cp})^{\top}$ by
 \begin{equation}
 \nonumber
  u_{cj}^{\ast}
  \begin{cases}
    \sim N(0,100) & \text{if $j$-th variable is assigned to $c$-th cluster},\\
    =0 & \text{otherwise},
  \end{cases}
    \quad j=1,\ldots,p.
 \end{equation}
Next, we generated a true task-specific parameter for $m$-th task 
that belongs to $c$-th cluster $\bm{v}^{c\ast}_{m} = (v^{c\ast}_{m1},\ldots,v^{c\ast}_{mp})^{\top}$ by
 \begin{equation}
 \nonumber
  v_{mj}^{c\ast}
  \begin{cases}
    \sim N(0,1)  & \text{if $j$-th variable is assigned to $c$-th cluster},\\
    =0 & \text{otherwise},
  \end{cases}
      \quad j=1,\ldots,p.
 \end{equation}
For Case 1, we generated a true outlier parameter for $m$-th task belonging to $c$-th cluster but assigned to an outlier task $\bm{o}_{m}^{c\ast}=(o_{m1}^{c\ast},\ldots,o_{mp}^{c\ast})^{\top}$ by 
\begin{equation}
\nonumber
  o_{mj}^{c\ast}
  \begin{cases}
    \sim f^{\mathrm{MTN}}(o)
& \text{if $j$-th variable is assigned to $c$-th cluster},\\
    =0 & \text{otherwise},
  \end{cases}
      \quad j=1,\ldots,p,
 \end{equation}
 where $f^{\mathrm{MTN}}(o)$ is a mixture of truncated normal distribution given by
 \begin{equation}
 \nonumber
    f^{\mathrm{MTN}}(o)=0.5f(o,-\infty,-3,-3,\sigma_{o}^{2})+ 0.5f(o,3,\infty,3,\sigma_{o}^{2}),
 \end{equation}
where $f(o,a,b,\mu,\sigma)$ is a truncated normal distribution on $a\leq o \leq b$  whose original normal distribution has mean $\mu$ and variance $\sigma^{2}$.  The reason for generating $o_{mj}^{(c)\ast}$ from $f^{\mathrm{MTM}}(o)$ is to leave the absolute value of $\bm{o}_{mj}^{c\ast}$ away from zero so that the outlier task is located away from the cluster. For Case 2, we generated a true outlier parameter for $m$-th task $\bm{o}_{m}^{\ast}=(o_{m1}^{\ast},\ldots,o_{mp}^{\ast})^{\top}$ by
\begin{equation}
\nonumber
    o_{mj}^{\ast} \sim U(-10,10), \quad  j=1,\ldots,p,
\end{equation}
where $U(a,b)$ is the continuous uniform distribution.

From these generating ways of $\bm{u}_{c}^{\ast},\bm{v}_{m}^{c\ast}$, and $\bm{o}_{m}^{\ast}$,
true regression coefficient vectors for non-outlier tasks that belong to different clusters have different non-zero variables. In other words, tasks belonging to different clusters are orthogonal to each other.
This orthogonal setting has been used in some studies of MTL (\cite{Jacob2008-ji}; \cite{Zhou2016-ik}).

 For our true model, we set  as $n_{m}=200$, $p=100$, $T=150$, and $\sigma^{2}=5$. $200$ samples in each task were split into $50$ samples for the train, $100$ samples for the validation, and left samples for the test.  We considered settings: $\kappa=\{0,0.1,0.2,0.3,0.4\}$.

We compared MTLRRC with several methods for the evaluation. For MTLRRC, we consider the three cases where group lasso (GL), group SCAD (GS), and group MCP (GM) are used for the fourth term in \eqref{MTLRRC}. Here, we set $\gamma=3.7$ for group SCAD and $\gamma=3$ for group MCP. 
As other competing methods, we employed MTLCVX, MTLK (multi-task learning via $k$-means; \cite{Argyriou2007-cj}), and Hotelling-like outlier task detection with MTLK (HMTLK). Here, HMTLK was done as follows. First, $\widehat{\bm{w}}_{m}^{\mathrm{STL}}\; (m=1,\ldots,T)$ were estimated and these sample mean $\Bar{\bm{w}}^{\mathrm{STL}}$ and covariance matrix $\Bar{\Sigma}^{\mathrm{STL}}$ were also calculated.
 For these values, we calculated the statistic
$h_{m}=(\widehat{\bm{w}}_{m}^{\mathrm{STL}}-\Bar{\bm{w}}^{\mathrm{STL}})^{\top}(\Bar{\Sigma}^{\mathrm{STL}})^{-1}(\widehat{\bm{w}}_{m}^{\mathrm{STL}}-\Bar{\bm{w}}^{\mathrm{STL}})$.
 Then, we detected tasks that satisfied $h_{m}\geq \chi^{(95)}_{p}$, where $\chi^{(95)}_{p}$ is a $95$ percentile point of $\chi^{2}$ distribution having $p$ degrees of freedom. Finally, we estimated regression coefficients by MTLK except for detected outlier tasks. Note that the detection based on Hotelling's $T^{2}$ is not theoretically justified for this simulation setting. However, as we will see later, it is possible to detect some outlier tasks.
 
 The weights $r_{m_{1},m_{2}}$ for both MTLCVX and MTLRRC were calculated by Eq. (\ref{setR}). $k$ was set to five. The estimation of $\widehat{\bm{w}}_{m}^{\mathrm{STL}}$ in Eq. (\ref{setR}) and HMTLK were performed by the lasso in R package ``glmnet''. The tuning parameter $\nu$ included in Algorithm \ref{MTLRRC_ADMM} was set to one. The regularization parameters were determined by the validation data. For the evaluation, 
we calculated the normalized mean squared error (NMSE), root mean squared error (RMSE),  true positive rate (TPR), and false positive rate (FPR):
    \begin{equation}
    \nonumber
        \begin{split}
        \mathrm{NMSE} &= \frac{1}{T} \sum_{m=1}^{T}\frac{\|\bm{y}^{\ast}_{m}-X_{m}\widehat{\bm{w}}_{m} \|_{2}^{2}}{n_{m}\mathrm{Var}(\bm{y_{m}^{\ast}})},\\
        \mathrm{RMSE} &= \frac{1}{T}\sqrt{\sum_{m=1}^{T}\|\bm{w}^{\ast}_{m}-\widehat{\bm{w}}_{m} \|_{2}^{2}},\\
         \mathrm{TPR} &= \frac{\#\left\{m;\bm{o}_{m}^{\ast} \neq\bm{0} \;\land\;\widehat{\bm{o}}_{m}\neq\bm{0}\right\}}{\#\left\{m;\bm{o}_{m}^{\ast}\neq \bm{0}\right\}}, \\
        \mathrm{FPR} &= \frac{\#\left\{m;\bm{o}_{m}^{\ast} = \bm{0} \;\land \;\widehat{\bm{o}}_{m}\neq\bm{0}\right\}}{\#\left\{m;\bm{o}_{m}^{\ast}= \bm{0}\right\}}.
        \end{split}
    \end{equation}
NMSE and RMSE evaluate the accuracy of the prediction and estimated regression coefficients, respectively. TPR and FPR evaluate the accuracy of outlier detection. They were computed $40$ times, and the mean and standard deviation were obtained in each setting.

Tables \ref{sim_result_case1} and \ref{sim_result_case2} show the results of simulation studies for Cases 1 and 2, respectively. For Case 1, MTLRRC and MTLCVX show almost the same accuracy in NMSE and RMSE with all $\kappa$s. MTLRRC, MTLCVX, and MTLK, which do not remove the outlier task a priori, outperform HMTLK in terms of NMSE and RMSE. This may suggest that multi-task learning improves the estimation accuracy even for outlier tasks that are located away from other tasks. As for TPR, HMTLK shows better results than MTLRRC. If the purpose is only to detect and eliminate outlier tasks, HMTLK is probably a better choice than robust MTL methods. For FPR, MTLRRC with non-convex penalties archives almost the best performance with any $\kappa$.  For Case 2, MTLRRC with non-convex penalties shows slightly better performance than MTLCVX in terms of NMSE and RMSE. Furthermore, MTLRRC with non-convex penalties is superior to HMTLK in terms of both TPR and FPR.

On the whole, MTLRRC with non-convex penalties detected true outlier tasks while greatly minimizing the detection of false outlier tasks for both Case 1 and Case 2. However, the differences in estimation accuracy between MTLCVX and MTLRCC are small, particularly for small $\kappa$. Thus, if the observed tasks consist of underlying clusters and a few outlier tasks, it would be difficult to improve the estimation and prediction accuracy by MTLRRC.
On the other hand, MTLRRC with the group lasso regularization shows poor performance even for TPR and FPR. For outlier task detection, the group lasso regularization would be not recommended.

\begin{table}[!ht]
     \caption{Simulation result of Case\;1}
     \label{sim_result_case1}
    \begin{tabular}{clcccc}
    \hline
$\kappa$ & Method &NMSE & RMSE & TPR & FPR   \\ \hline
0      & MTLRRC (GL) & $\bm{0.010}$ (0.003)  & 0.424 (0.073) & N/A & 0.206 (0.358) \\ 
      & MTLRRC (GS) & $\bm{0.010}$ (0.004) & $\bm{0.417}$ (0.073) & N/A &  $\bm{0.001}$ (0.002)    \\ 
      & MTLRRC (GM) & 0.011 (0.003) & 0.424 (0.076) & N/A & $\bm{0.001}$ (0.003)    \\ 
      & HMTLK & 0.080 (0.061) & 1.302 (0.469) & N/A & 0.09 (0.08)    \\ 
      & MTLCVX & $\bm{0.010}$ (0.003) & 0.427 (0.076) & -- & --    \\ 
      & MTLK & 0.029 (0.044) & 0.603 (0.377) & -- & --    \\ \hline
0.1 & MTLRRC (GL) & $\bm{0.035}$ (0.008) & 1.056 (0.111) & 0.386 (0.443) & 0.263 (0.414) \\ 
      & MTLRRC (GS) & $\bm{0.035}$ (0.007) & 1.043 (0.120) & 0.345 (0.434) & $\bm{0.001}$ (0.002)    \\ 
      & MTLRRC (GM) & 0.036 (0.008)	 & 1.048 (0.120) & 0.363 (0.416) & $\bm{0.001}$ (0.004)     \\ 
      & HMTLK & 0.112 (0.039) & 1.864 (0.314) & $\bm{0.724}$ (0.152) & 0.073 (0.061)    \\ 
      & MTLCVX & $\bm{0.035}$(0.009) & $\bm{1.030}$ (0.131) & -- & --   \\ 
      & MTLK & 0.054 (0.040)	 & 1.190 (0.274) & -- & --    \\ \hline
0.2 & MTLRRC (GL) & 0.063 (0.012) & 1.459 (0.136) & 0.249 (0.356) & 0.094 (0.279) \\ 
      & MTLRRC (GS) & 0.069 (0.015) & 1.490 (0.146) & 0.261 (0.381) & $\bm{0.001}$ (0.003)     \\ 
      & MTLRRC (GM) & 0.066 (0.012) & 1.459 (0.154) & 0.352 (0.427) & $\bm{0.001}$ (0.004)    \\ 
      & HMTLK & 0.145 (0.040)	 & 2.234 (0.290) &  $\bm{0.693}$ (0.080) & 0.040 (0.049)    \\ 
      & MTLCVX & $\bm{0.062}$ (0.014) & $\bm{1.437}$ (0.145) & -- & --    \\ 
      & MTLK & 0.079 (0.044) & 1.528 (0.258) & -- & --    \\ \hline
0,3 & MTLRRC (GL) & 0.092 (0.019) & 1.778 (0.140) & 0.381 (0.426) & 0.200 (0.359) \\ 
      & MTLRRC (GS) & 0.092 (0.020) & $\bm{1.761}$ (0.154) & 0.149 (0.255) & $\bm{0.001}$ (0.003)    \\ 
      & MTLRRC (GM) & 0.095 (0.017) & 1.823 (0.143) & 0.453 (0.440)	 & $\bm{0.001}$ (0.004)     \\
      & HMTLK & 0.189 (0.045)	 & 2.596 (0.233) & $\bm{0.622}$ (0.078) & 0.023 (0.025)    \\
      & MTLCVX & $\bm{0.091}$ (0.013) & 1.778 (0.113) & -- & -- \\
      & MTLK & 0.106 (0.049)	 & 1.836 (0.298) & -- & --    \\ \hline
0,4 & MTLRRC (GL) & 0.120 (0.020) & 2.043 (0.122) & 0.306 (0.382) & 0.179 (0.333) \\ 
      & MTLRRC (GS) & $\bm{0.117}$ (0.020) & 2.035 (0.154) & 0.317 (0.394) & $\bm{0.001}$ (0.003)    \\ 
      & MTLRRC (GM) & $\bm{0.117}$ (0.017) & $\bm{2.026}$ (0.117) & 0.313 (0.383) & 0.002 (0.004)    \\ 
      & HMTLK & 0.244 (0.044) & 2.929 (0.196) & $\bm{0.586}$ (0.092) & 0.036 (0.049)    \\ 
      & MTLCVX & 0.122(0.020) & 2.089 (0.139) & -- & --    \\ 
      & MTLK & 0.135 (0.056) & 2.090 (0.288) & -- & --    \\ \hline
    \end{tabular}
\end{table}

\begin{table}[!ht]
     \caption{Simulation result of Case\;2}
     \label{sim_result_case2}
     \begin{tabular}{clcccc}
     \hline
$\kappa$&Method&NMSE & RMSE & TPR & FPR   \\ \hline
    0.1 &MTLRRC (GL) & 0.069 (0.015) & 1.205 (0.141) & 0.294 (0.337) & 0.304 (0.446) \\ 
	&MTLRRC (GS) & 0.069 (0.015) & 1.202 (0.130) & $\bm{0.886}$ (0.308) & 0.014 (0.013)    \\ 
	&MTLRRC (GM) & $\bm{0.067}$ (0.014) & $\bm{1.188}$ (0.132) & 0.826 (0.358) & $\bm{0.008}$ (0.008)  \\ 
	&HMTLK & 0.157 (0.060) & 1.867 (0.388) & 0.613 (0.134) & 0.111 (0.078)  \\ 
	&MTLCVX & 0.072 (0.017) & 1.233 (0.163) &--&-- \\ 
	&MTLK & 0.094 (0.035) & 1.380 (0.236) &--&-- \\ \hline
    0.2 &MTLRRC (GL) & 0.130 (0.019) & 1.670 (0.135) & 0.480 (0.361) & 0.339 (0.414) \\ 
	&MTLRRC (GS) & 0.125 (0.021) & 1.631 (0.136) & $\bm{0.936}$ (0.215)  & $\bm{0.020}$ (0.019)  \\ 
	&MTLRRC (GM) & $\bm{0.118}$ (0.016) & $\bm{1.595}$ (0.115) & 0.935 (0.224) & 0.023 (0.013)   \\ 
	&HMTLK & 0.211 (0.044) & 2.155 (0.251) & 0.524 (0.100)	 & 0.075 (0.066)  \\ 
        &MTLCVX & 0.123 (0.020)	 & 1.625 (0.139) &--&-- \\ 
	&MTLK & 0.152 (0.035) & 1.793 (0.201)  &--&--\\ \hline
    0.3 &MTLRRC (GL) & 0.173 (0.017) & 1.935 (0.103) & 0.467 (0.352) & 0.317 (0.387) \\ 	
        &MTLRRC (GS) & 0.171 (0.018) & 1.914 (0.102) & $\bm{0.936}$ (0.226) & $\bm{0.020}$ (0.020)  \\ 
	&MTLRRC (GM) & $\bm{0.166}$ (0.018) & $\bm{1.881}$ (0.114) & 0.911 (0.246) & 0.021 (0.020)  \\ 
	&HMTLK & 0.276 (0.042) & 2.459 (0.198) & 0.542 (0.089) & 0.066 (0.051)   \\ 
	&MTLCVX & 0.181 (0.024) & 1.972 (0.132)  &--&--\\ 
	&MTLK & 0.204 (0.037) & 2.081 (0.179) &--&--\\ \hline
    0.4 &MTLRRC (GL) & 0.232 (0.022) & 2.246 (0.109) & 0.551 (0.387) & 0.292 (0.391) \\ 
	&MTLRRC (GS) & $\bm{0.220}$ (0.024) & $\bm{2.181}$ (0.115) & $\bm{0.943}$ (0.202) & $\bm{0.029}$ (0.025)  \\ 
	&MTLRRC (GM) & 0.226 (0.028) & 2.210 (0.134) & 0.890 (0.275) & 0.031 (0.023) \\ 
	&HMTLK & 0.337 (0.049) & 2.698 (0.199) & 0.482 (0.078) & 0.047 (0.043)  \\ 
	&MTLCVX & 0.232 (0.024) & 2.243 (0.122) &--&-- \\ 
        &MTLK & 0.285 (0.047) & 2.471 (0.196)&--&-- \\ \hline
\end{tabular}
\end{table}

\section{Application to real datasets}
\label{Sec6}
In this section, we apply MTLRRC to two real datasets.
The first is the landmine data \citep{Xue2007-wl}, which consists of nine-dimensional features and the corresponding binary labels for 29 tasks. Each task corresponds to a landmine field where data were collected: 1--15 tasks
correspond to regions that are relatively highly foliated and 16--29 tasks correspond to regions that are bare earth or desert. Thus, there may be two clusters depending on the ground surface conditions. The responses represent landmines or clutter. The features are four moment-based features, three correlation-based features, one energy ratio feature, and one spatial variance feature. These are extracted from radar images. This dataset contains 14,820 samples in total, but the number of positive samples and negative samples is quite unbalanced. To perform the analysis stably, down-sampling was done by reducing negative samples to equal the number of positive samples. In the results, we used 1,808 samples in total.
The second dataset is the school data \citep{Bakker2003-tp}, which consists of examination scores of 15,362 students, four school-specific attributes, and three student-specific attributes from 139 secondary schools in London from 1985 to 1987. The dataset was obtained by the ``MALSAR'' package in MATLAB. In the package, categorical attributes were replaced with binary attributes. Then, we used 28-dimensional explanatory variables and the examination scores as a response. Each school is considered a task.

\begin{table}[htbp]
    \centering
    \caption{AUC and NMSE for landmine data and school data in 100 repetitions}
    \label{result_realdata}
    \begin{tabular}{lcc}
    \hline
        ~ & Landmine & School  \\ 
        Method & AUC  & NMSE   \\ \hline
        MTLRRC\,(GS) &  $\bm{0.764}$ (0.021) & 0.853 (0.058)  \\ 
        MTLRRC\,(GS$\gamma$) & $\bm{0.764}$ (0.024) & $\bm{0.844}$ (0.049)\\
        MTLRRC\,(GM) & 0.760 (0.026) & 0.852 (0.053) \\ 
        MTLRRC\,(GM$\gamma$) & 0.761 (0.026) & 0.852 (0.058) \\
        MTLCVX & 0.760 (0.022) & 0.847 (0.048) \\ 
        MTLK & 0.756 (0.023) & 0.847 (0.048) \\ \hline
    \end{tabular}
\end{table}

\begin{figure}[htbp] 
    \begin{tabular}{cc}
    \multicolumn{2}{c}{
    \begin{minipage}[t]{0.5\hsize}
        \centering
        \includegraphics[width=6cm]{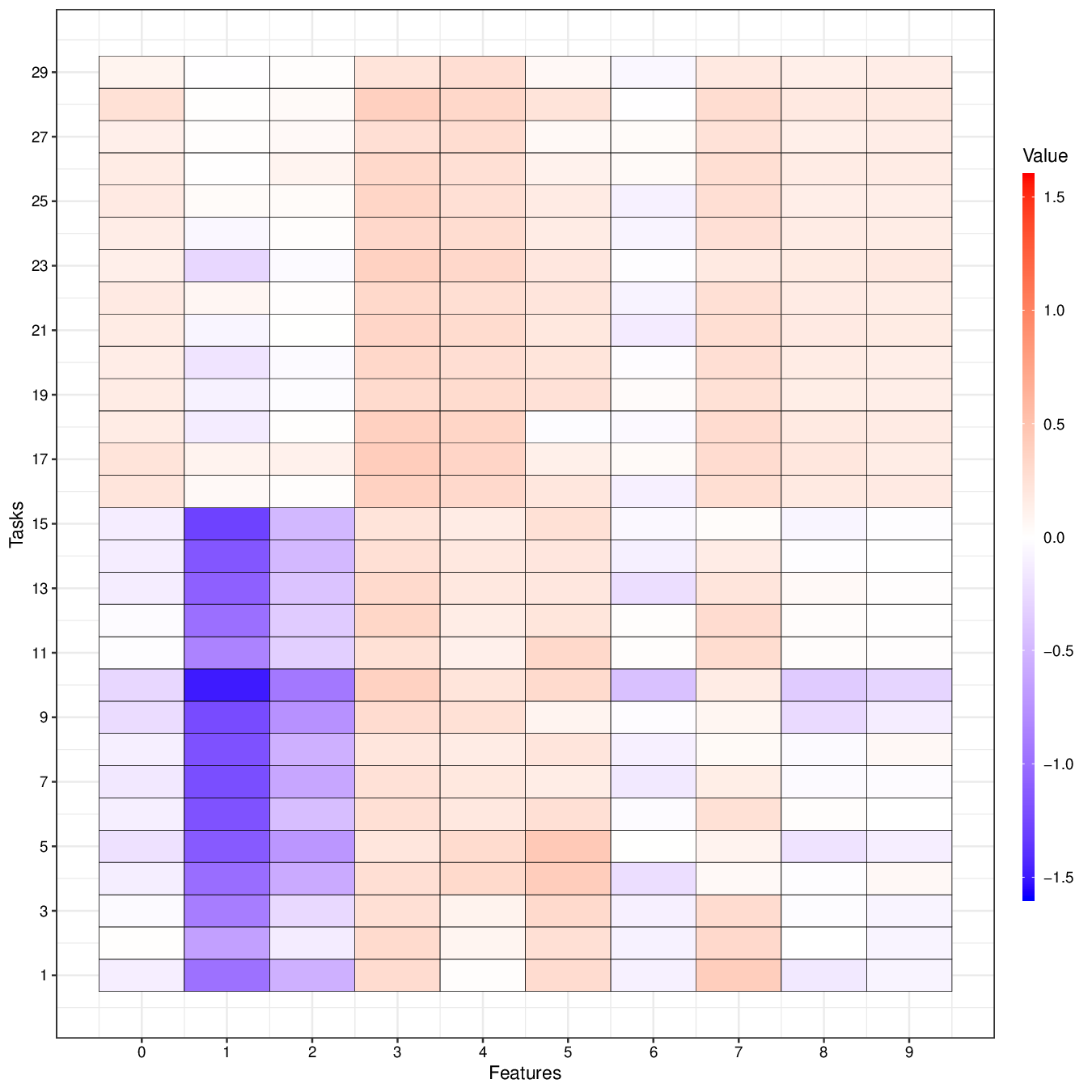}
        \subcaption{Mean of estimated $\widehat{W}$ for 100 repetitions in the landmine data}
        \label{lanemine_meanW}
    \end{minipage}}
    \\
    \begin{minipage}[t]{0.5\hsize}
        \centering
        \includegraphics[width=6cm]{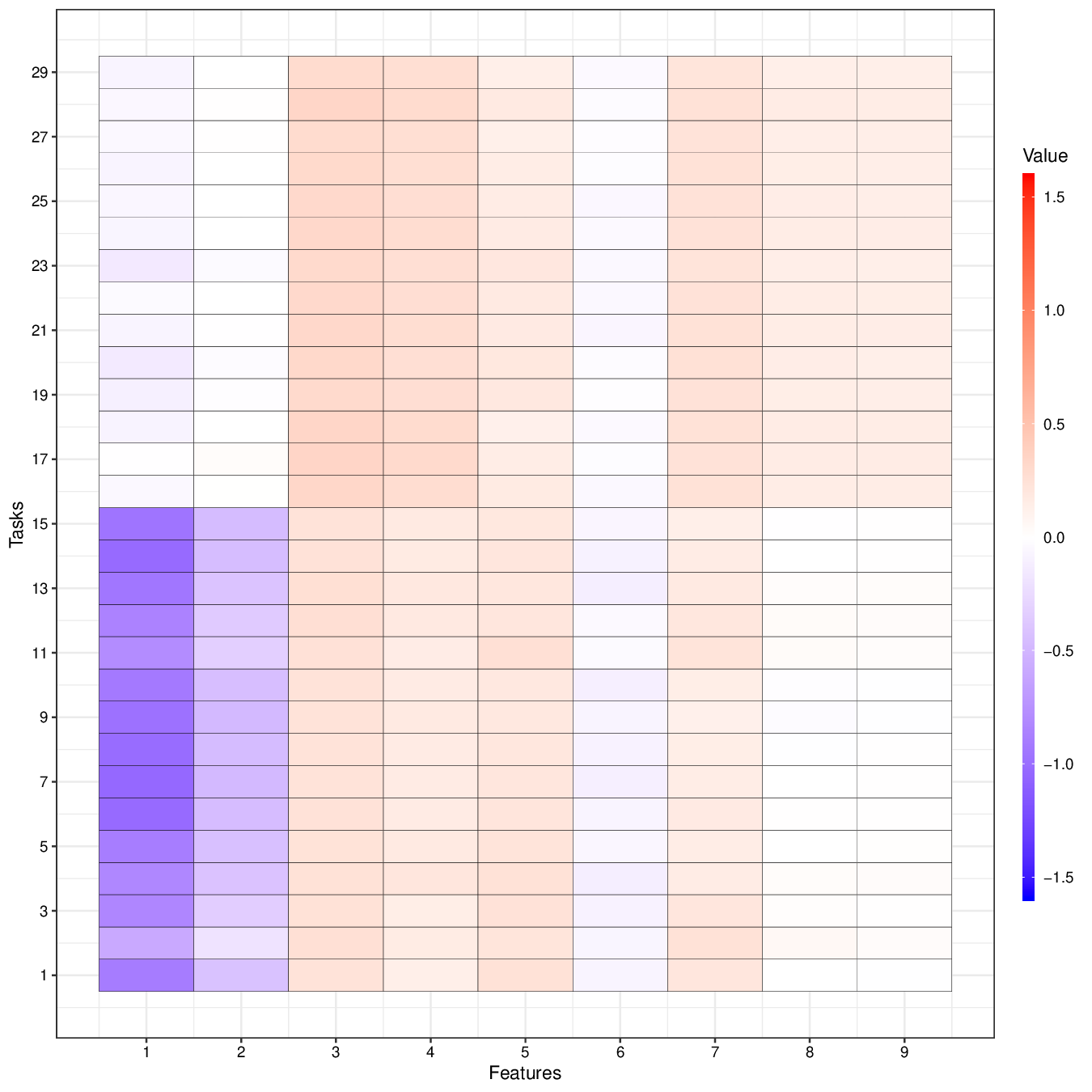}
        \subcaption{Mean of estimated $\widehat{U}$ for 100 repetitions in the landmine data}
        \label{lanemine_meanU}
      \end{minipage}
      &
      \begin{minipage}[t]{0.5\hsize}
        \centering
        \includegraphics[width=6cm]{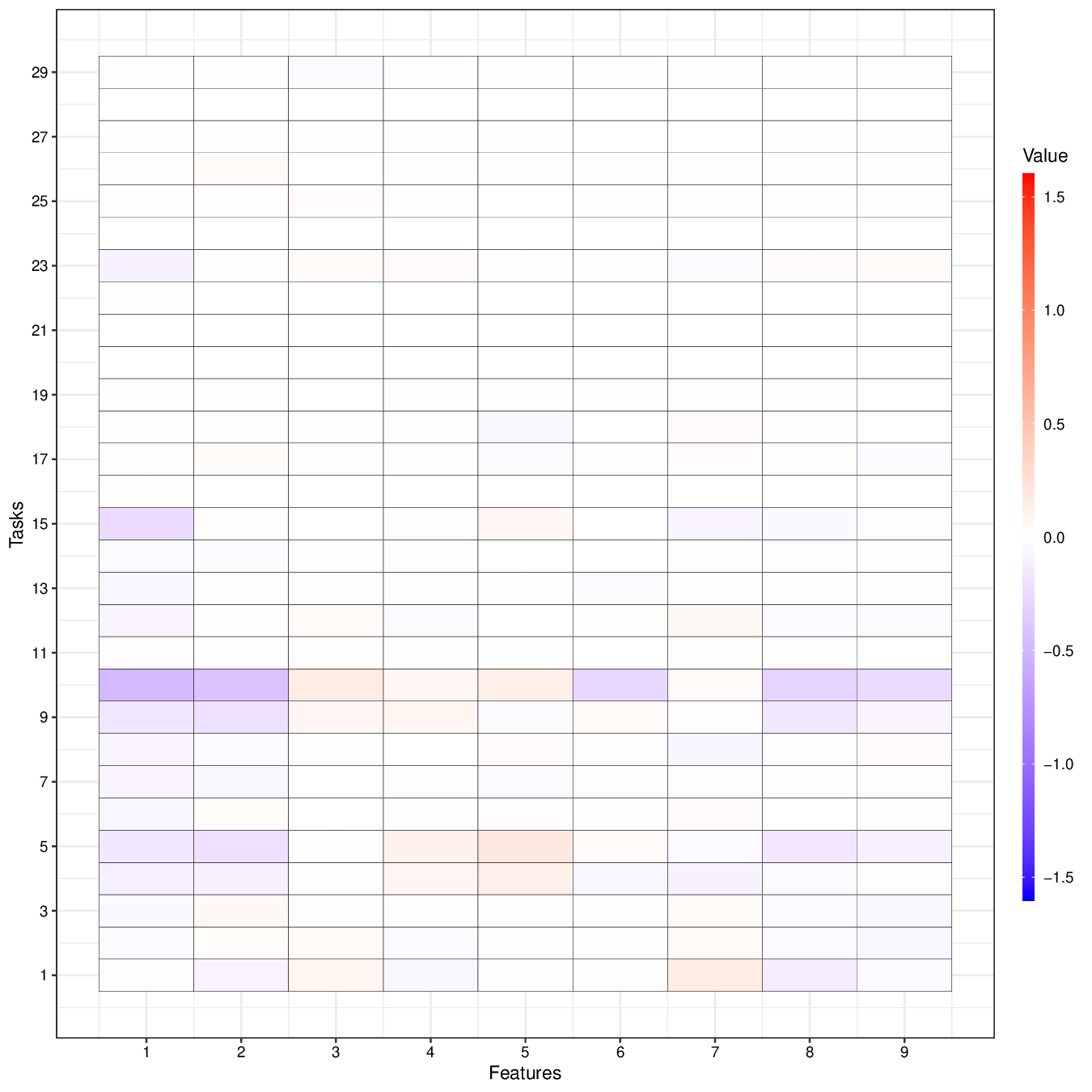}
        \subcaption{Mean of estimated $\widehat{O}$ for 100 repetitions in the landmine data}
        \label{lanemine_meanO}
      \end{minipage}
    \end{tabular}
    \caption{The mean of the estimated value of parameters in MTLRRC\,(GS$\gamma$) in 100 repetitions for the landmine data}
    \label{landmine_parameter}
  \end{figure}
\begin{figure}[htbp]
    \begin{tabular}{cc}
    \multicolumn{2}{c}{
    \begin{minipage}[t]{0.5\hsize}
        \centering
        \includegraphics[width=6cm]{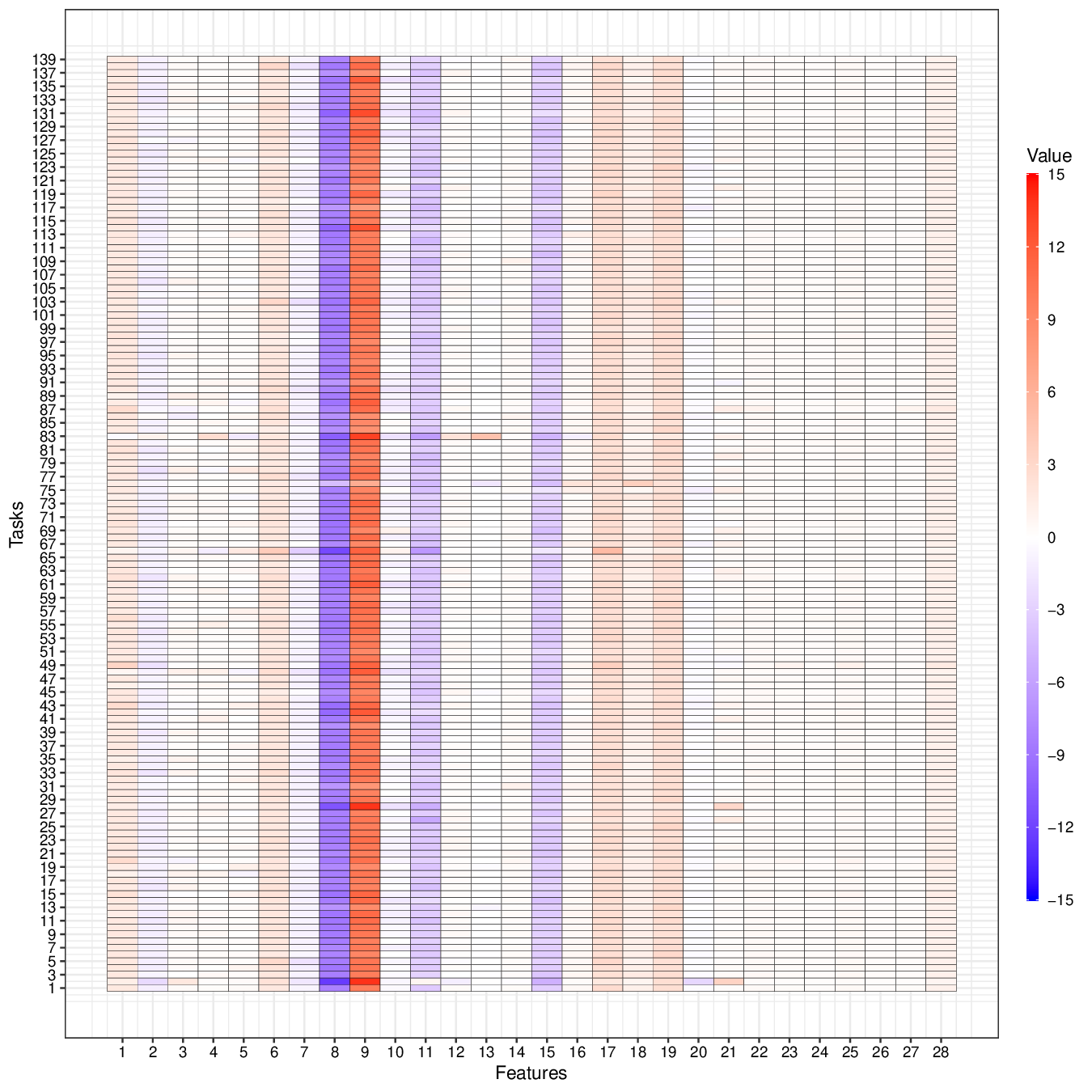}
        \subcaption{Mean of estimated $\widehat{W}$ for 100 repetitions in the school data}
        \label{school_meanW}
    \end{minipage}}
    \\
    \begin{minipage}[t]{0.5\hsize}
        \centering
        \includegraphics[width=6cm]{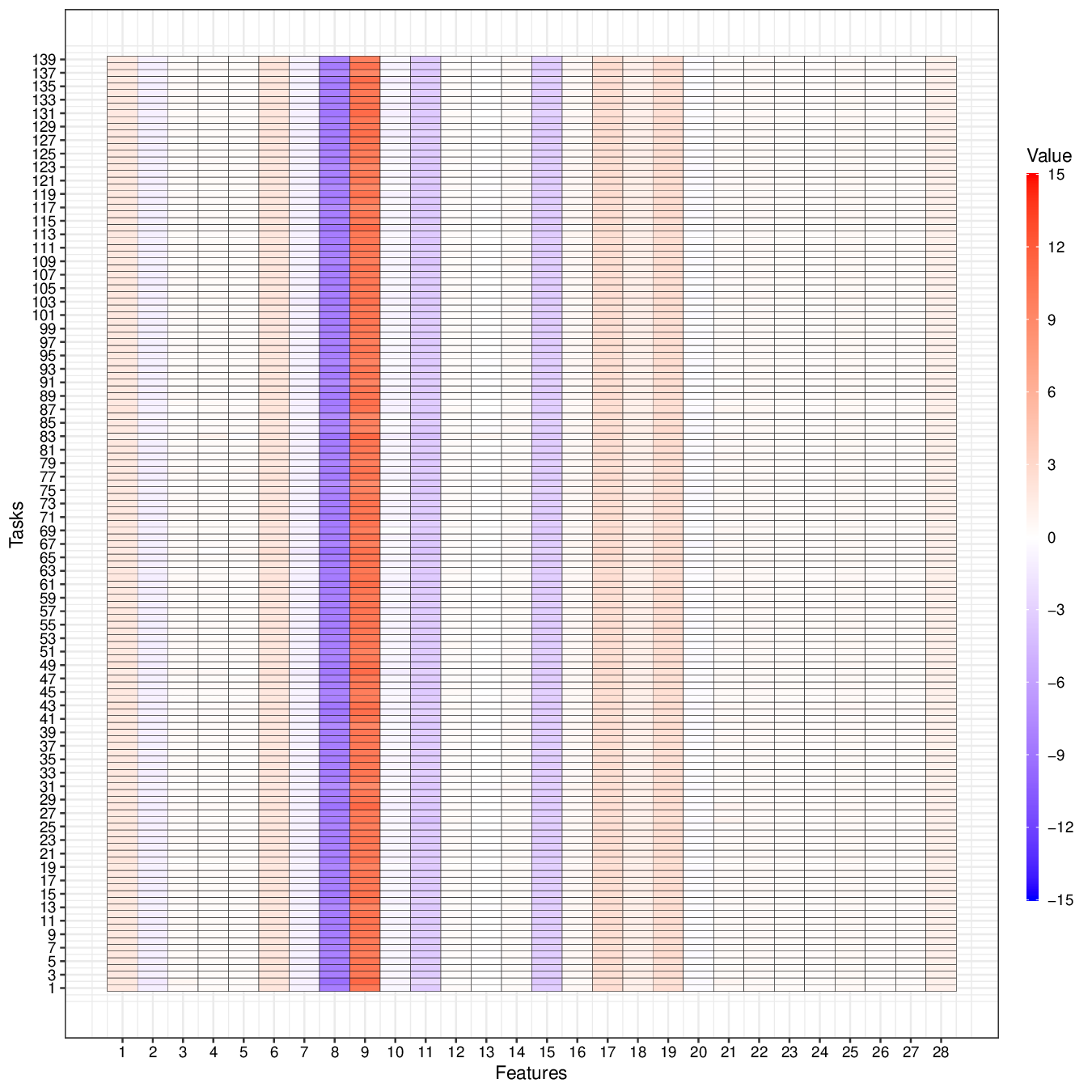}
        \subcaption{Mean of estimated $\widehat{U}$ for 100 repetitions in the school data}
        \label{school_meanU}
      \end{minipage}
      &
    \begin{minipage}[t]{0.5\hsize}
        \centering
        \includegraphics[width=6cm]{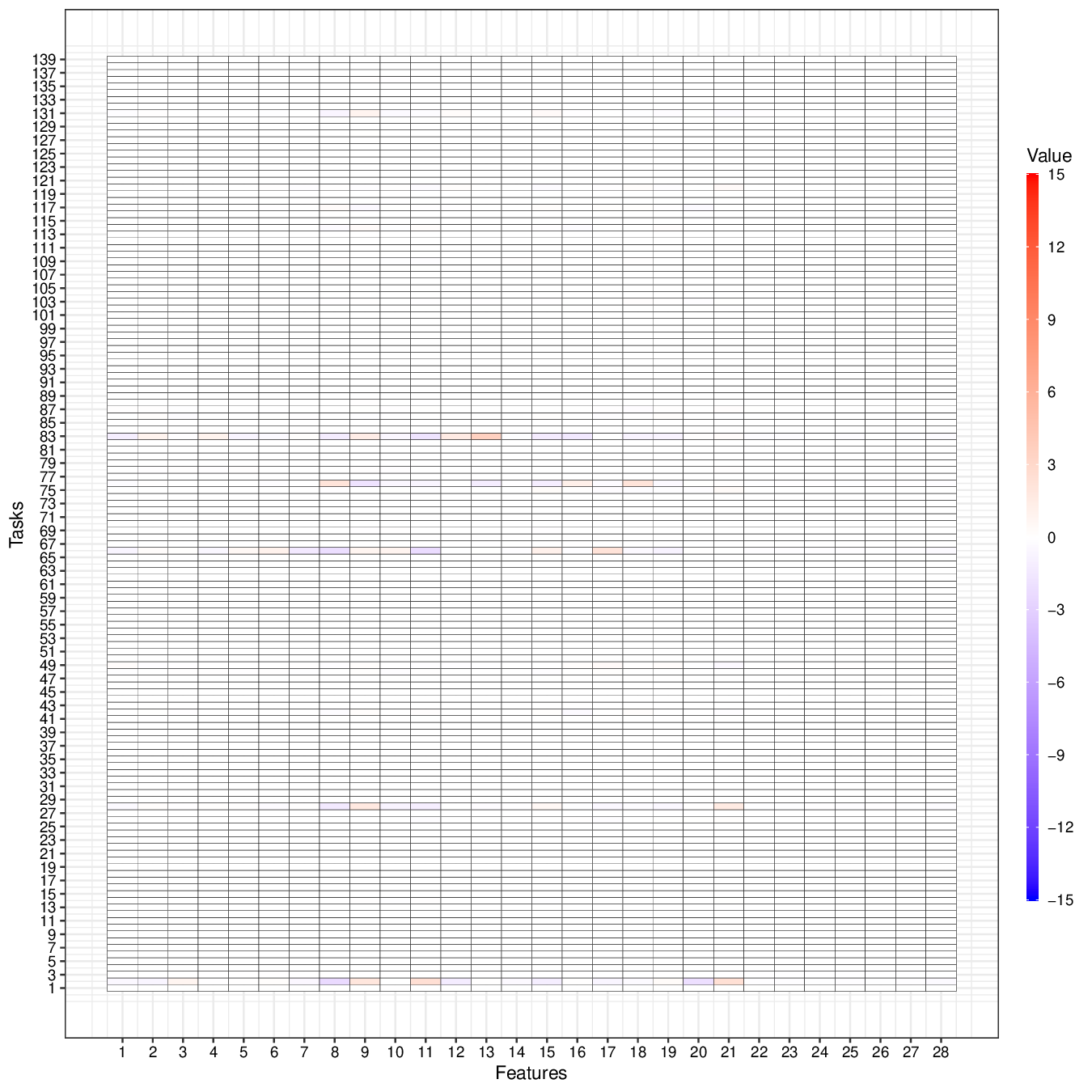}
        \subcaption{Mean of estimated $\widehat{O}$ for 100 repetitions in the school data}
        \label{school_meanO}
      \end{minipage}
    \end{tabular}
    \caption{The mean of the estimated value of parameters in MTLRRC\,(GS$\gamma$) in 100 repetitions for the school data}
    \label{school_parameters}
\end{figure}

\begin{figure}[htbp]
 \centering
        \includegraphics[width=9cm]{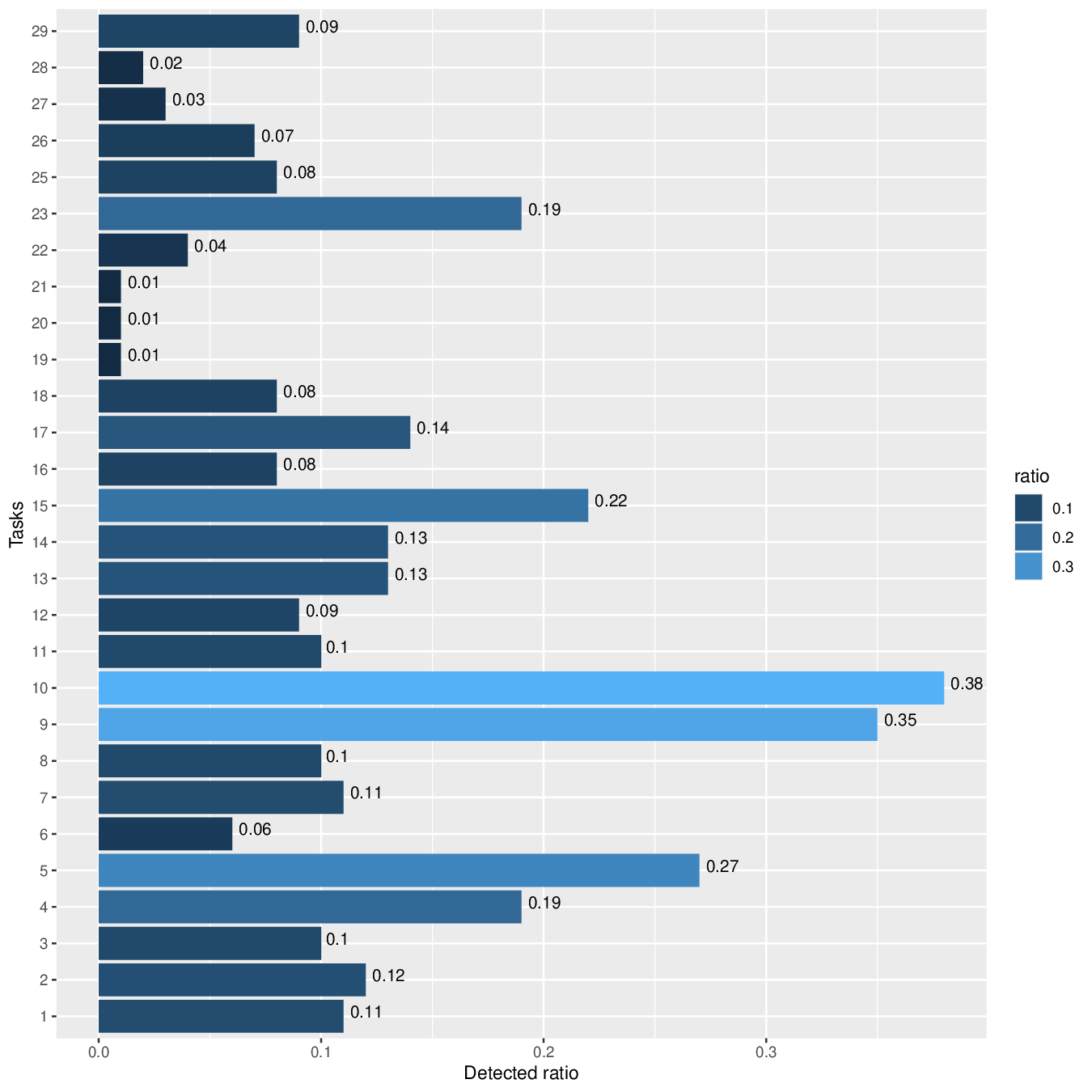}
        \caption{Ratio of $\widehat{\bm{o}
}_{m}\neq\bm{0}$ for 100 repetitions in the landmine data}
        \label{landmine_outlier_prob}
\end{figure}
\begin{figure}[htbp]
 \centering
        \includegraphics[width=9cm]{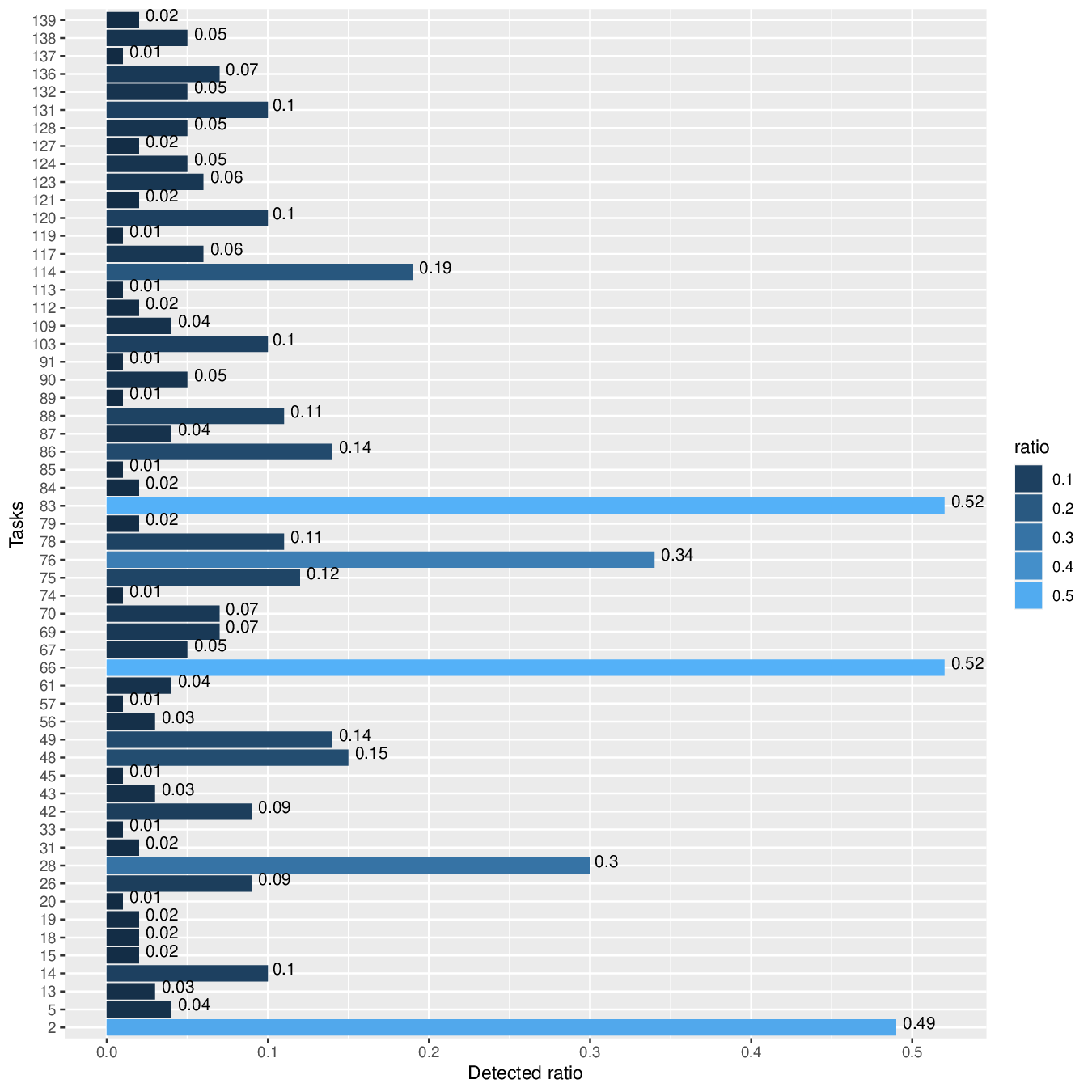}
        \caption{Ratio of $\widehat{\bm{o}
}_{m}\neq\bm{0}$ for 100 repetitions in the school data}
        \label{school_outlier_prob}
\end{figure}

We compared MTLRRC with MTLCVX and MTLK. First, we split the samples in each task into $60\%$ for the train, $20\%$ for the validation, and $20\%$ for the test. The regularization parameters are determined by the validation data. In MTLRRC, we use non-convex penalties and consider four cases. In the first case, $\gamma$ is fixed with $3$ for the group MCP (GM). In the second case, $\gamma$ is fixed with $3.7$ for the group SCAD (GS). In the third and fourth cases, $\gamma$ is chosen by the validation data for group MCP and group SCAD, respectively (GM$\gamma$ and GS$\gamma$). For the evaluation, we calculated NMSE and AUC for school data and landmine data, respectively. These values were calculated 100 times with the random splitting of the dataset. 

Table \ref{result_realdata} shows the mean and standard deviation in 100 repetitions. From this table, we observe that MTLRRC\,(GS$\gamma$) gives a smaller NMSE and larger AUC than other methods for the school data and the landmine data, respectively.

Next, we calculated the mean of estimated parameters $\widehat{W},\widehat{U}$, and $\widehat{O}$ in MTLRRC\,(GS$\gamma$). 
 Figures \ref{landmine_parameter} and \ref{school_parameters} show the mean of estimated parameter values for the landmine data and the school data, respectively. The vertical axis is the index of tasks and the horizontal axis is the index of features. Each color shows the mean of the estimated parameters. Figures \ref{landmine_outlier_prob} and \ref{school_outlier_prob} are bar plots that show the ratios of each task detected as an outlier task. Here, note that tasks that have never been detected are removed from the bar plots. 
 
 For the landmine data, Figure \ref{landmine_parameter}(\subref{lanemine_meanU}) suggests the presence of two clusters, which is consistent with the fact that tasks 1--15 and 16--29 are obtained from regions corresponding to different surface conditions. For outliers detection in Figure \ref{landmine_parameter}(\subref{lanemine_meanO}), because the $10$-th task has a relatively larger or smaller value than other tasks and the task is a task detected as an outlier with a ratio greater than 0.3 from Figure \ref{landmine_outlier_prob}, it may indicate that the task is a potential outlier task. Furthermore, the $9$-th task also has a relatively high detection ratio that is greater than 0.3, although the mean of $\widehat{\bm{o}}_{m}$ is not as clear as the $10$-th task from Figure \ref{landmine_parameter}(\subref{lanemine_meanO}).
 The estimated regression coefficients for the $9$-th and $10$-th tasks in Figure \ref{landmine_parameter}(\subref{lanemine_meanW}) show a similarity to other tasks within the same estimated cluster. These results may suggest the underlying structure among tasks in the landmine data is rather Case 1 than Case 2 in Section \ref{Sec5}. Furthermore, from Figure \ref{landmine_outlier_prob}, we observe that only two tasks in tasks 16--29 have an outlier task ratio of more than 0.1, while 13 tasks in tasks 1--15 have. This result suggests that the cluster composed of tasks 1--15 may have relatively large variability. This may provide some insight into the structure concerning sub-groups within the cluster.

 For the school data, we obtained the homogeneous pattern in Figure \ref{school_parameters}(\subref{school_meanU}). The school data have been considered rather homogeneous in some studies (\cite{Bakker2003-tp};  \cite{Evgeniou2005-it}). Thus, this result would be reasonable and shows that MTLRRC can consistently estimate cluster structure even when the true number of underlying clusters is one. From Figure \ref{school_outlier_prob}, the 2nd, 28th, 66th, 76th, and 83rd tasks were detected with a rate greater than 0.3. In addition, these tasks show larger values in $\bm{o}_{m}$ than other tasks from Figure \ref{school_parameters}(\subref{school_meanO}). For the regression coefficients in Figure \ref{school_parameters}(\subref{school_meanW}), although these tasks share many characteristics with other tasks, some regression coefficients have different characteristics. For example, the regression coefficients for $76$th and $83$rd tasks of $16$th feature have, respectively, relatively large and small values that are much different from other tasks.

\section{Conclusion}
\label{Sec7}

In this paper, we proposed a robust multi-task learning method called Multi-Task Learning via Robust Regularized Clustering (MTLRRC). To perform the clustering of tasks and detection of outlier tasks simultaneously, we incorporated regularization terms based on robust regularized clustering (RRC), which can detect outlier samples by selecting outlier parameters through group sparse penalties. We showed that the solution of the RRC obtained by the BCD algorithm shares the same first-order condition with convex clustering whose loss function is replaced by the multivariate robust loss function. Thus, MTLRRC is expected to perform robust clustering of tasks. Furthermore, the solution of MTLRRC by the BCD algorithm is also viewed as a convergence point of alternative optimization that involves the RRC for tasks and regression problems reducing the shrinkage of outlier tasks toward the estimated centroid. To mitigate computational costs, we developed an estimation algorithm based on the modified ADMM.

We conducted simulation studies to evaluate the performance of MTLRRC under two scenarios of outlier task structures. In Case 1, outlier tasks share the same characteristics as the centroid but have additional outlier parameters. In Case 2, outlier tasks do not share any common structure and are independent of other tasks. 
 In both of these cases, the proposed method with non-convex group penalties exhibited a near-zero FNR in outlier detection and a much larger TPR in Case 2. However, the improvement in estimation and prediction accuracy was slight, especially when the proportion of outlier tasks was few.

In the application to real data, we observed that the proposed method effectively estimates multiple cluster structures and identifies potential outlier tasks resembling Case 1. These findings suggest that MTLRRC not only estimates clusters but also provides insights into the heterogeneity of outlier tasks within clusters.

One limitation of our study is that MTLRRC includes three or four regularization parameters to be determined. The computational cost of searching for optimal value of these parameters can be demanding. On the other hand, the definition of the multivariate \textit{M}-estimator having a connection to the group penalties is different from that of the traditional one \citep{Maronna1976-mh}: while the former \textit{M}-estimator is defined as a straightforward extension of the univariate \textit{M}-estimator with robust multivariate loss function, the traditional one is defined as the solution of weighted log-likelihood equations. Although there may be some relationship between these two definitions, they are probably not equivalent. Moreover, outlier tasks similar to Case 2 were not detected in the analyzed real data. We leave these topics as future work.

\backmatter

\bmhead{Acknowledgments}
A. O. was supported by JST SPRING, Grant Number JPMJSP2136.
S. K. was supported by JSPS KAKENHI Grant Numbers JP23K11008, JP23H03352, and JP23H00809. 
Supercomputing resources were provided by the Human Genome Center (the Univ. of Tokyo).


\begin{appendices}
\section{Multivariate robust loss functions}
    \label{app_loss}
    In this appendix, we show some multivariate loss functions and corresponding group-thresholding functions.
    \subsection{Group SCAD}
    For $\gamma>2$, the group SCAD thresholding function and loss function are, respectively, expressed as
    \begin{equation}
    \nonumber
    \rho_{\lambda,\gamma}^{\mathrm{gSCAD}}(\bm{z})=
        \begin{cases}
            \frac{1}{2} \|\bm{z}\|_{2}^{2} & \|\bm{z}\|_{2} \leq \lambda,\\
            \lambda \|\bm{z}\|_{2}  - \frac{\lambda^{2}}{2} & \lambda \leq \|\bm{z}\|_{2} < 2\lambda, \\
            \frac{\gamma\lambda}{\gamma-2}\|\bm{z}\|_{2}-\frac{1}{2(\gamma-2)}\|\bm{z}\|_{2}^{2}-\frac{\gamma+2}{2(\gamma-2)}\lambda^{2} &  2\lambda \leq \|\bm{z}\|_{2} \leq \gamma\lambda, \\
            \frac{\gamma+1}{2}\lambda^{2} & \gamma\lambda < \|\bm{z}\|_{2}, \end{cases}
    \end{equation}
    \begin{equation}
    \nonumber
    \bm{\Theta}^{\mathrm{gSCAD}}(\bm{o};\lambda,\gamma)=
    \begin{cases}
    S(\bm{o};\lambda), & \|\bm{o}\|_{2}\leq 2\lambda,\\
    \frac{\gamma-1}{\gamma-2}S(\bm{o},\frac{\gamma\lambda}{\gamma-1}), & 2\lambda<\|\bm{o}\|_{2}\leq\gamma\lambda,\\
    \bm{o}, & \|\bm{o}\|_{2}>\gamma\lambda.
    \end{cases}
    \end{equation}
    
    \subsection{Group MCP}
    For $\gamma>1$, the group MCP thresholding function and loss function are, respectively, expressed as
     \begin{equation}
     \nonumber
    \rho_{\lambda,\gamma}^{\mathrm{gMCP}}(\bm{z})=
        \begin{cases}
            \frac{1}{2} \|\bm{z}\|_{2}^{2} & \|\bm{z}\|_{2} \leq \lambda,\\
            \frac{\gamma\lambda}{\gamma-1}\|\bm{z}\|_{2}-\frac{1}{2(\gamma-1)}\|\bm{z}\|_{2}^{2}-\frac{\gamma\lambda^{2}}{2(\gamma-1)} &  \lambda \leq \|\bm{z}\|_{2} \leq \gamma\lambda, \\ \frac{\gamma\lambda^{2}}{2} & \gamma\lambda < \|\bm{z}\|_{2},           \end{cases}
    \end{equation}
    \begin{equation}
    \nonumber
    \bm{\Theta}^{\mathrm{gMCP}}(\bm{o};\lambda,\gamma)=
    \begin{cases}
        \frac{\gamma}{\gamma-1}S(\bm{o},\lambda), & \|\bm{o}\|_{2}\leq\gamma\lambda,\\
    \bm{o}, & \|\bm{o}\|_{2}>\gamma\lambda.
    \end{cases}
    \end{equation}
    
    \subsection{Multivariate Tukey}
    We define the multivariate version of Tukey's loss function as  
     \begin{equation}
     \nonumber
        \rho_{\lambda}^{\mathrm{MT}}(\bm{z}) =\begin{cases}
            1-\left(1-\frac{\|\bm{z}\|_{2}^{2}}{\lambda^{2}}\right)^{3} & \|\bm{z}\|_{2}\leq \lambda, \\
            1 & \|\bm{z}\|_{2} > \lambda.
            \end{cases}
    \end{equation}
    The corresponding group thresholding function can be expressed as 
    \begin{equation}
    \nonumber
        \Theta^{\mathrm{MT}}(\bm{o};\lambda,\gamma) =\begin{cases}
            \bm{o}-\bm{o}\left(1-\frac{\|\bm{o}\|_{2}^{2}}{\lambda^{2}}\right)^{2} & \|\bm{o}\|_{2} \leq\lambda, \\
            \bm{o} & \|\bm{o}\|_{2} > \lambda.
            \end{cases}
    \end{equation}
    
\section{Proofs}
\label{appendix:proofs}
\subsection{Proof of Proposition \ref{propo1}}
\label{appendix:proofs1}
 From the update of Algorithm \ref{BCD_GRRC}, a convergence point $(\widehat{U},\widehat{O})$ satisfies 
\begin{equation}
\nonumber
\mathrm{vec}(\widehat{O}) = \begin{pmatrix} \widehat{\bm{o}}_{1} \\
\widehat{\bm{o}}_{2} \\
\vdots \\
\widehat{\bm{o}}_{n}
\end{pmatrix}=\begin{pmatrix}
\bm \Theta(\bm{x}_{1}-\widehat{\bm{u}}_{1};\lambda_{2},\gamma) \\
\bm \Theta(\bm{x}_{2}-\widehat{\bm{u}}_{2};\lambda_{2},\gamma) \\
\vdots \\
\bm \Theta(\bm{x}_{n}-\widehat{\bm{u}}_{n};\lambda_{2},\gamma) \\
\end{pmatrix}.
\end{equation}
On the other hand, because the update \eqref{updata_U} is equivalently expressed as
\begin{equation}
\nonumber
\widehat{U} =
 \argmin_{\substack{U}}\left\{ \frac{1}{2}\|\mathrm{vec}(X)-\mathrm{vec}(U)-\mathrm{vec}(\widehat{O})\|_{2}^{2}+ \lambda_{1}\|D\mathrm{vec}(U)\|_{2,1}\right\},
\end{equation}
it follows:
\begin{equation}
\nonumber
 -\{\mathrm{vec}(X)-\mathrm{vec}(\widehat{U})-\mathrm{vec}(\widehat{O})\} + \lambda_{1} \frac{\partial}{\partial\mathrm{vec}(U)} \|D\mathrm{vec}(U)\|_{2,1}\rvert_{U=\widehat{U}} = \bm{0}.
\end{equation}
Thus, we conclude Proposition \ref{propo1} from
\begin{equation}
\nonumber
\begin{split}
   &\mathrm{vec}(X)-\mathrm{vec}(\widehat{U})-\mathrm{vec}(\widehat{O}) \\
   &=\begin{pmatrix}
    \bm{x}_{1}-\widehat{\bm{u}}_{1}-\bm \Theta(\bm{x}_{1}-\widehat{\bm{u}}_{1};\lambda_{2},\gamma) \\
    \bm{x}_{2}-\widehat{\bm{u}}_{2}-\bm \Theta(\bm{x}_{2}-\widehat{\bm{u}}_{2};\lambda_{2},\gamma) \\
    \vdots \\
    \bm{x}_{n}-\widehat{\bm{u}}_{n}-\bm \Theta(\bm{x}_{n}-\widehat{\bm{u}}_{n};\lambda_{2},\gamma) \\
    \end{pmatrix} \\
    & =\begin{pmatrix} \psi (\bm{x}_{1}-\bm{\widehat{u}}_{1};\lambda_{2},\gamma) \\ \psi (\bm{x}_{2}-\bm{\widehat{u}}_{2};\lambda_{2},\gamma) \\ \vdots \\ \psi (\bm{x}_{n}-\bm{\widehat{u}}_{n};\lambda_{2},\gamma) \end{pmatrix} \\
    &= \Psi(X-\widehat{U};\lambda_{2},\gamma).
 \end{split}
\end{equation}
\subsection{Proof of Proposition \ref{propo2}}
\label{appendix:proofs2}
From Algorithm \ref{BCD_MTLRRC},  a convergence point $(\widehat{\bm w}_0, \widehat{W},\widehat{U},\widehat{O})$ satisfies
\begin{equation}
\label{propo2_update_O}
\mathrm{vec}(\widehat{O}) = \begin{pmatrix} \widehat{\bm{o}}_{1} \\
\widehat{\bm{o}}_{2} \\
\vdots \\
\widehat{\bm{o}}_{T}
\end{pmatrix}=\begin{pmatrix}
\bm \Theta(\widehat{\bm{w}}_{1}-\widehat{\bm{u}}_{1};\lambda_{3}/\lambda_{1},\gamma) \\
\bm \Theta(\widehat{\bm{w}}_{2}-\widehat{\bm{u}}_{2};\lambda_{3}/\lambda_{1},\gamma) \\
\vdots \\
\bm \Theta(\widehat{\bm{w}}_{T}-\widehat{\bm{u}}_{T};\lambda_{3}/\lambda_{1},\gamma) \\
\end{pmatrix},
\end{equation}
\begin{equation}
\label{propo2_update_W}
    (\widehat{w}_{m0},\widehat{\bm{w}}_{m}) = \argmin_{w_{m0},\bm{w}_{m}} \left\{\frac{1}{n_{m}} L(w_{m0},\bm{w}_{m})+ \frac{\lambda_{1}}{2} \|\bm{w}_{m}-\widehat{\bm{u}}_{m}-\widehat{\bm{o}
}_{m}\|_{2}^{2} \right\}, \quad m=1,\ldots,T,
\end{equation}
\begin{equation}
\label{propo2_update_U}
    \widehat{U} =
    \argmin_{\substack{U}}\left\{ \frac{\lambda_{1}}{2}\|\mathrm{vec}(\widehat{W})-\mathrm{vec}(U)-\mathrm{vec}(\widehat{O})\|_{2}^{2}+ \lambda_{2}\|D\mathrm{vec}(U)\|_{2,1}\right\}.
\end{equation}
From the first-order condition of the minimization problems \eqref{propo2_update_W} and \eqref{propo2_update_U}, we obtain
\begin{align} 
\nonumber
\frac{\partial}{\partial \bm{w}_{m}^{\prime}}\frac{1}{n_{m}}L(w_{m0},\bm{w}_{m})\rvert_{\bm{w}_{m}^{\prime}=\widehat{\bm{w}}_{m}^{\prime}} + \lambda_{1} \begin{pmatrix}
0 \\
\widehat{\bm{w}}_{m}-\widehat{\bm{u}}_{m}-\widehat{\bm{o}}_{m}
\end{pmatrix} &= \bm{0}, \\ \nonumber
    - \lambda_{1}\{ \mathrm{vec}(\widehat{W})-\mathrm{vec}(\widehat{U})-\mathrm{vec}(\widehat{O}) \} + \lambda_{2} \frac{\partial}{\partial\mathrm{vec}(U)} \|D\mathrm{vec}(U)\|_{2,1}\rvert_{U=\widehat{U}} &= \bm{0}.
\end{align}
Some algebra for them and \eqref{propo2_update_O} conclude Proposition \ref{propo2}.

\section{Derivation of update for $W, B$, and $S$ in Algorithm \ref{MTLRRC_ADMM}}
\label{appendix_ADMM}
To illustrate the derivation of update for $W, B$, and $S$, we show the idea of \cite{Shimmura2022-ar}.

Let $A$ be a $p \times n$ matrix, and let $f:\mathbb{R}^{n}\rightarrow\mathbb{R}$ and $h:\mathbb{R}^{m}\rightarrow\mathbb{R}$ be convex functions. We consider the minimization problem taking form:
\begin{equation}
\nonumber
     \min_{\bm{x},\bm{y}}\left\{ f(\bm{x})+h(\bm{y})\right\}, \quad \mathrm{s.t.} \quad A\bm{x}=\bm{y}.
\end{equation}
Then, the augmented Lagrangian is expressed as
\begin{equation}
\nonumber
    L_{\nu}(\bm{x},\bm{y},\bm{s})= f(\bm{x}) + h(\bm{y}) + \bm{s}^{\top}(A\bm{x}-\bm{y}) + \frac{\nu}{2}\|A\bm{x}-\bm{y}\|_{2}^{2},
\end{equation}
where $\bm s$ is a $p$-dimensional vector of Lagrangian multipliers and $\nu \ (\geq 0)$ is a tuning parameter. 
For this augmented Lagrangian, 
the jointly update of $\bm{x}$ and $\bm{y}$ based on ADMM is expressed as 
\begin{equation}
\nonumber
    \begin{split}
        (\bm{x}^{(t+1)},\bm{y}^{(t+1)}) = \argmin_{\bm x,\bm y}\left\{ L_{\nu}(\bm{x},\bm{y},\bm{s}^{(t)}) \right\}.
    \end{split}
\end{equation}
If we consider updating only $\bm{x}$ by the joint optimization, the update can be written as
\begin{equation}
        \label{update_x}
    \begin{split}
        \bm{x}^{(t+1)}     &= \argmin_{\bm{x}}\left\{\min_{\bm{y}}L_{\nu}(\bm{x},\bm{y},\bm{s}^{(t)}) \right\} \\
        &= \argmin_{\bm{x}}\left\{ f(\bm x)+ \min_{\bm{y}} \left\{ h(\bm{y})+\bm{s}^{\top(t)}(A\bm{x}-\bm{y}) + \frac{\nu}{2}\|A\bm{x}-\bm{y}\|_{2}^{2}\right\} \right\} \\
        & = \argmin_{\bm{x}}\left\{f(\bm{x}) - r^{\ast}(\nu A\bm{x}+s^{(t)}) + \bm{s}^{\top(t)}A\bm{x}+\frac{\nu}{2}\|A\bm{x}\|_{2}^{2}\right\},
    \end{split}
\end{equation}
where $r^{\ast}(\cdot)$ is the conjugate function of $r(\bm{z})=h(\bm{z})+\frac{\nu}{2}\|\bm{z}\|_{2}^{2}$. 
This indicates that we can update $\bm{x}$ without updating $\bm{y}$. Furthermore, the following lemma motivates us to solve the minimization problem \eqref{update_x} by the gradient method.

\begin{lemma}[\cite{Shimmura2022-ar}; Theorem 1]
If we define $\phi_{1}(\bm x)=f(\bm x)+\min_{\bm{y}}(h(\bm{y})+\bm{s}^{\top(t)}(A\bm{x}-\bm{y}) + \frac{\nu}{2}\|A\bm{x}-\bm{y}\|_{2}^{2})$, $\phi_{1}$ is differentiable. Furthermore, we have
    \begin{equation}
    \nonumber
        \frac{\partial}{\partial \bm{x}}\phi_{1}(\bm{x}) = \frac{\partial}{\partial \bm{x}} f(\bm{x})+A^{\top}(\mathrm{prox}_{\nu h^{\ast}}(\nu A \bm{x}+\bm{s}^{(t)})),
    \end{equation}
    where $\mathrm{prox}_{g}(\cdot)$ is the proximal map of $g(\cdot)$ defined as 
    \begin{equation}
    \nonumber
        \mathrm{prox}_{g}(\bm{z}) = \argmin_{\bm{x}}\left\{g(\bm{x}) +\frac{1}{2}\|\bm{z}-\bm{x} \|_{2}^{2} \right\}.
    \end{equation}
\end{lemma}
As a result, the update of $\bm{x}$ is given by the convergence point of the gradient method. Here, the update of the gradient method is given by 
\begin{equation}
\nonumber
    \bm{x}^{(l+1)} = \bm{x}^{(l)} - \iota\left\{\frac{\partial}{\partial \bm{x}} f(\bm{x})\rvert_{\bm{x}=\bm{x}^{(l)}}+A^{\top}(\mathrm{prox}_{\nu h^{\ast}}(\nu A \bm{x}^{(l)}+\bm{s}^{(t)}))\right\},
\end{equation}
where $\iota$ is a step size with a non-negative value. \cite{Shimmura2022-ar} considered accelerating the convergence rate of the gradient method based on fast iterative shrinkage-thresholding algorithm (FISTA; \cite{Beck2009-qe}). To guarantee the convergence, FISTA requires setting $\iota\leq 1/L$ for $L>0$ such that 
\begin{equation}
\nonumber
    \left\|\frac{\partial}{\partial \bm{x}} \phi_{1}(\bm{x})\rvert_{\bm{x}=\bm{x}_{1}}-\frac{\partial}{\partial \bm{x}} \phi_{1}(\bm{x})\rvert_{\bm{x}=\bm{x}_{2}} \right\|_{2}^{2} \leq L\|\bm{x}_{1}-\bm{x}_{2} \|_{2}^{2}.
\end{equation}
The update of $\bm{s}$ via ADMM can be converted by Moreau decomposition as
\begin{equation}
    \label{update_s}
    \begin{split}
        \bm{s}^{(t+1)} &= \bm{s}^{(t)}+\nu(A\bm{x}^{(t+1)}-\bm{y}^{(t+1)})  \\
        &= \mathrm{prox}_{\nu h^{\ast}}(\bm{s}^{(t)}+\nu A \bm{x}^{(t+1)}).
    \end{split}
\end{equation}

By applying these ideas to the augmented Lagrangian of MTLRRC in \eqref{augmented_lagrangian_MTLRRC}, we obtain the update concerning the gradient method as
\begin{equation}
    \label{update_U_gradient}
    U^{(l+1)} = U^{(l)}-\iota\{\lambda_{1}(U^{(l)}+O^{(t)}-W^{(t+1)})+A_{\mathcal{E}}^{\top}F\},
\end{equation}
where $F$ is the matrix whose each row is updated by
\begin{equation}
\nonumber
    F_{(m_{1},m_{2})}=\mathrm{prox}((S^{(t)}+ \nu A_{\mathcal{E}} \cdot C^{(l)})_{(m_{1},m_{2})},\lambda_{2}r_{m_{1},m_{2}}).
\end{equation}
Here, $\mathrm{prox}(\cdot,\lambda)$ defined in 
 Eq. \eqref{prox_group} is a proximal map of $h^{\ast}$ derived from $h(\bm z)=\lambda\|\bm{z}\|_{2}$. As a result, by combining the update \eqref{update_U_gradient} and FISTA, we obtain an update for $U^{(t+1)}$ in Algorithm \ref{MTLRRC_ADMM}.
 
 Similarly, by applying update \eqref{update_s} to MTLRRC, the update of $S$ is given by 
 \begin{equation}
 \nonumber
     S^{(t+1)}_{(m_{1},m_{2})} = \mathrm{prox}((S^{(t)}+\nu A_{\mathcal{E}} \cdot U^{(t+1)})_{(m_{1},m_{2})},\lambda_{2}r_{m_{1},m_{2}}).
 \end{equation}

\end{appendices}

\noindent

\bigskip




\bibliography{MTLRRC.bib}



\end{document}